\documentclass{jfm}
\usepackage{graphicx}
\usepackage{epstopdf, epsfig}
\usepackage{xcolor}
\usepackage{changes}

\usepackage{comment}
\usepackage{color}
\usepackage{subfigure}

\shorttitle{Guidelines for authors}
\shortauthor{X.J. Wang, M.P. Wan and L. Biferale}

\title{Acceleration of tracer and light particles
in compressible homogeneous isotropic turbulence}

\author{Xiangjun Wang\aff{1,2},
  and Minping Wan\aff{1,3}
  \corresp{\email{wanmp@sustech.edu.cn}}
  and Luca Biferale\aff{4}}

\affiliation{\aff{1}Guangdong Provincial Key Laboratory of Turbulence Research and Applications, Department of Mechanics and Aerospace Engineering,Southern University of Science and Technology, Nanshan District, Shenzhen, Guangdong, 518055, China
\aff{2}Harbin Institute of Technology, Nangang District, Harbin, Heilongjiang, 150090, China
\aff{3}Southern Marine Science and Engineering Guangdong Laboratory (Guangzhou), 1119 Haibin Road, Nansha District, Guangzhou 511458, China
\aff{4}Department of Physics and INFN University of Rome `Tor Vergata', Via della Ricerca Scientifica 1, 00133 Rome, Italy}

\begin{document}

\maketitle

\begin{abstract}
The accelerations of tracer and light particles
in compressible homogeneous isotropic turbulence (CHIT)
is investigated by using data from  direct numerical simulations (DNS)
up to turbulent  Mach number $M_t =1$.
For tracer particles, the flatness factor of
acceleration components, $F_a$, increases gradually
for $M_t \in [0.3, 1]$. On the contrary, $F_a$ for light particles
 develops a maximum around  $M_t \sim 0.6$.
The PDF of longitudinal acceleration of tracers
is increasingly skewed towards the negative value as $M_t$ increases.
By contrast, for light particles,
the skewness factor of longitudinal acceleration, $S_a$,
firstly becomes more negative with the increase of $M_t$,
and then goes back to $0$ when $M_t$ is larger than $0.6$.
Similarly, differences among tracers and light particles appear also in the zero-crossing time of acceleration correlation.
It is argued that all these phenomenons are intimately linked to
the flow structures in compression regions, e.g. close to shocklets.
\end{abstract}

\begin{keywords}
Authors should not enter keywords on the manuscript, as these must be chosen by the author during the online submission process )
\end{keywords}

\section{Introduction}
Much attention has been paid to the investigation
of particle-laden turbulence in the past decades.
The behaviors of tracer and inertial particles in incompressible turbulence
have been studied extensively,  as for diffusion,
collision and  preferential concentration
\citep{jullien1999richardson, ott2000experimental,
boffetta2002relative, boffetta2002statistics,
wang1998collision, zhou1998collision, yamamoto2001large,
sommerfeld2001validation, vance2006properties,
onishi2014collision, maxey1987gravitational,
falkovich2002acceleration, wilkinson2006caustic,
chun2005clustering, goto2008sweep, coleman2009unified,
bragg2015mechanisms, bragg2015relationship}.
However, there have been much less study
of compressible particle-laden turbulence,
which are important to understand
flow in different domains as, e.g., scramjets
{\citep{ferri1973mixing,curran1996fluid,
huete2016weak,urzay2018supersonic}} and the motion of interstellar gas
\citep{barth2007trams,johansen2007rapid,pan2011turbulent}.
In this paper, we will focus mainly on the acceleration statistics
of tracer and light particles dispersed in Compressible Homogeneous and Isotropic Turbulence  (CHIT), an idealized set-up
which could be helpful to the investigation on cavitating flow
with clouds of bubbles \citep{reisman1998observations,
wang1999numerical,fuster2011modelling}.

Investigation of particle acceleration in turbulence
has attracted much attention in the last 20 years.
Following the ``K41'' theory \citep{kolmogorov1941dissipation,
kolmogorov1941local}, the Heisenberg-Yaglom model
was proposed to predict the variance of fluid acceleration
in homogeneous isotropic turbulence (HIT)
\citep{heisenberg1985statistischen, yaglom1949acceleration}:
$$\langle \textbf{\emph{a}}^2 \rangle = a_0 \varepsilon^{3/2} \nu^{-1/2}.$$
Here $\textbf{\emph{a}}$ is the acceleration vector,
$\langle \cdot \rangle$ represents the ensemble average,
$a_0$ is a universal constant, which was further found
to have an anomalous dependency on the Taylor microscale Reynolds number,
$R_{\lambda}$, owing to the turbulent intermittency
\citep{yeung1989lagrangian,voth1998lagrangian,vedula1999similarity,
gotoh1999intermittency,biferale2004multifractal}.
$\varepsilon=2\nu \langle S_{ij}S_{ij} \rangle$
is the average dissipation rate,
$S_{ij}$ is the rate-of-strain tensor,
and $\nu$ is the kinematic viscosity.
\citet{yeung1989lagrangian} numerically found that in HIT  for fluid particles,
the zero-crossing time ($\tau_0$, the time in which
the autocorrelation function deceases to 0)
of acceleration components is around $2.2$ times Kolmogorov timescale
($\tau_{\eta}$).
Moreover, the ratio of $\tau_0$ to $\tau_{\eta}$
barely varies with Reynolds number, which is supported
by later investigations \citep{yeung1997one, voth1998lagrangian,
mordant2004three, biferale2005joint, volk2008acceleration}.
Subsequently, \citet{yeung1997one} discovered
that $\tau_0$ of the acceleration magnitude
is much larger than $\tau_{\eta}$.
According to the momentum equation,
the contribution to fluid acceleration
can be decomposed into two parts:
pressure gradient term and viscous force term.
\citet{vedula1999similarity} pointed out that
the contribution of pressure gradient term is dominant over
that of viscous force in turbulence.
\citet{la2001fluid} experimentally confirmed that
the acceleration of fluid particles is extremely intermittent
in fully developed turbulence.
The probability density function (PDF) of acceleration components
is well reproduced by a superposition of stretched exponentials
\citep{biferale2004multifractal} and the flatness factor of any component,
$F_a = \langle (a_i)^4 \rangle/ \langle (a_i)^2 \rangle^2 $, with $i=1,2,3$,
can exceed 60 when $R_{\lambda}$ is up to 500.
{Later on, many investigations on tracer particle acceleration
in incompressible flows have been carried out
\citep{voth2002measurement,mordant2004three,xu2007acceleration,
lavezzo2010role,ni2012lagrangian,stelzenmuller2017lagrangian}.}

Apart from tracer particles, the acceleration of inertial particles
(both heavy and light particles) in incompressible turbulent flows
also attracted much attention.
\citet{bec2006acceleration} systematically studied
the acceleration statistics of heavy particles in fully developed HIT.
They pointed out that the flatness factor and the root-mean square value
of  particle accelerations both sharply decrease
with the increase of the Stokes number, $\mathrm{St} = \tau_p/\tau_\eta$,
given by the ratio of particle response time and the Kolmogorov flow time.
This is a combined effect of preferential concentration on
 regions with small pressure gradient and filtering due to inertia.
Later on, \citet{ayyalasomayajula2006lagrangian,ayyalasomayajula2008modeling}
proposed a vortex model to describe the selective sampling
of turbulent flows and filtering effects.
On the contrary, it is known that in HIT light particles enjoys the opposite behaviour,
with a strong preferential concentration on high vorticity regions,
with a PDF of the acceleration even more intermittent than the tracer case
and - in some cases - with longer correlation times when normalized by $\tau_{\eta}$
\citep{volk2008acceleration,volk2008laser}.
Recently, \citet{zhang2019model} proposed a theoretical model
which successfully predicts the variance of light particles acceleration
with different density ratios when St is small
(e.g. less than 1.0), as well as the dependence of
drag force and inertia force on St.

In recent years, developments of experimental
and numerical techniques prompted a new interest
on Lagrangian properties in compressible flows too
\citep{yang2013acceleration, yang2014interactions,
yang2016intermittency,zhang2016preferential,
zhang2018single,dai2017direct,dai2018direct,
xiao2020eulerian}.
\citet{yang2013acceleration} investigated the accelerations
of tracer particles in CHIT.
They found shocklets substantially increase the probability
of extremely large-acceleration events.
Moreover, almost all the large-acceleration events
come from compression regions.
Besides, the autocorrelation function of acceleration components
near shocklets decreases much faster than that near vortices.
Subsequently, \citet{yang2014interactions} studied
the influence of shocklets on inertial particles at $M_t \approx 1.0$.
They discovered that the PDF of the longitudinal acceleration
of light particles in compression regions
has a much wider tail at the positive value
than that of tracer and heavy particles.
It results from the fact that light particles
in compression regions have a higher probability
to develop a velocity pointing upstream (the low pressure side),
compared with tracer and heavy particles.
Recently, \citet{haugen2021spectral} studied the clustering of heavy particles
in two different compressible isotropic flows: compressively forced turbulence
and solenoidally forced turbulence.
They found that the clustering of particles in compressively forced turbulence
peaks at two different St.
The first peak (at lower St) results from the contribution of shocklets
and the other is attributed to the centrifugal sling effect.

In this paper,  we present a systematic investigation
of light particles accelerations statistics at high Reynolds numbers
and at changing the degree of compressibility, i.e. the turbulent Mach number
and the results will be compared with those of tracer particles.
In particular, we are interested to characterize the role played
by the presence of shocklets at changing $M_t$
concerning both the single time PDF of bubbles acceleration
and its temporal correlation.
One of the main result, is the identification of a {\it critical} Mach number,
$M_t \sim 0.6$ where the statistics of light particles
strongly departs from the one of tracers.
The outline of the paper is as follow: the numerical details are introduced
in Section \ref{sec:simulationconfiguration}.
The more important results and analysis are described in Section
\ref{sec:resultsandanalysis}.
Finally, a brief conclusion and discussion will be given in Section
\ref{sec:ConclusionandDiscussion}.

\section{Simulation configuration}\label{sec:simulationconfiguration}
The three-dimensional CHIT was simulated
in a periodic cubic box with each side length of $2\pi$,
based on a hybrid scheme \citep{wang2010hybrid}
which combines the seventh-order weighted essentially non-oscillatory
(WENO) scheme and eighth-order compact central finite difference scheme.
The dimensionless governing equations of compressible turbulent flow are shown below:
\begin{equation}
 \label{Eq4:mass} \frac{\partial \rho_f}{\partial t}+\nabla \cdot \left(\rho_f \textbf{\emph{u}} \right) =0
\end{equation}
\begin{equation}
 \label{Eq4:momentum} \frac{\partial \rho_f \textbf{\emph{u}}}{\partial t}+\nabla\cdot\left(\rho_f \textbf{\emph{u}} \textbf{\emph{u}}\right)=-\nabla \left(\frac{p}{\gamma M^2}\right)+\frac{1}{\mathrm{Re}} \nabla\cdot \boldsymbol{\sigma} +\textbf{\emph{f}}
\end{equation}
\begin{equation}
 \label{Eq4:energy} \frac{\partial \mathcal{E}}{\partial t}+\nabla\cdot\left[\left(\mathcal{E}+\frac{p}{\gamma M^2}\right)\textbf{\emph{u}} \right] =\frac{1}{\alpha}\nabla\cdot\left(\kappa \nabla T\right)+\frac{1}{\mathrm{Re}}\nabla\cdot\left(\boldsymbol{\sigma}\cdot \textbf{\emph{u}}\right) +\textbf{\emph{f}}\cdot\textbf{\emph{u}}+\Lambda
\end{equation}
\begin{equation}
 \label{Eq4:gas_state} p=\rho_f T
\end{equation}
where $\rho_f$, $\textbf{\emph{u}}$ and $p$ separately represent the density, velocity and static pressure of fluid. $\gamma$ is the ratio of the specific heat at constant pressure, $C_p$, to the specific heat at constant volume, $C_v$. $\boldsymbol{\sigma}=\mu\left[\nabla \textbf{\emph{u}} + (\nabla \textbf{\emph{u}})^T\right]-\frac{2}{3}\mu\theta\mathbf{I}$ is the viscous stress, where $\mu$ is the dynamic viscosity, $\theta=\nabla\cdot\textbf{\emph{u}}$ is the divergence of fluid velocity and $\mathbf{I}$ is the identity matrix.
Also, $M=U_0\slash c_0$ and $\mathrm{Re}=\rho_0U_0L_0\slash\mu_0$ are the reference Mach number and Reynolds number. Here, $L_0$, $\rho_0$, $U_0$, $\mu_0$ and $c_0$ are the reference length, density, velocity, dynamic viscosity and speed of sound separately. $\mathcal{E}=p\big/\left[(\gamma-1)\gamma M^2\right]+\frac{1}{2}\rho_f\textbf{\emph{u}}\cdot\textbf{\emph{u}}$ is the total energy per unit volume. $T$ denotes the temperature and $\kappa$ is the thermal conductivity.
$\alpha=\mathrm{Pr}\mathrm{Re} (\gamma-1)M^2$ is a coefficient from nondimensionalization, and $\mathrm{Pr}$ is the reference Prandtl number, which is specified as 0.7 in this paper.
To maintain a stationary flow, an external force ($\textbf{\emph{f}}$) is applied on the solenoidal velocity components at large scales
(the first two wavenumbers).
Besides, a uniform cooling term ($\Lambda$) is used to keep the internal energy steady.
{The detailed description about the forcing and cooling mechanism
was introduced in \cite{wang2010hybrid}, \cite{federrath2010comparing}
and \cite{kida1990energy}.}
In this paper, the turbulent Mach number ($M_t$) varies from 0.36 to 1.00
with Taylor Reynolds number $R_{\lambda} \approx 120$
and the total grid number ($N^3$) in the computational domain is $512^3$.
Detailed parameters of the simulations are listed in Table \ref{tab:param_f}.
\begin{table}
  \begin{center}
\def~{\hphantom{0}}
  \begin{tabular}{cccccccccc}
    $R_{\lambda}$&$M_t$&$N^3$&$\nu$&$\varepsilon$&$\eta$&$\tau_{\eta}$&$u_{rms}$&$L_{f}$&$T_{f}$\\[3pt]
    ~123~&~$0.36$~&~$512^3$~&~4.17E-3~&~0.73~&~1.78E-2~&~7.59E-2~&~$2.28$~&~1.47~&~1.12~\\
    ~122~&~$0.51$~&~$512^3$~&~4.18E-3~&~0.78~&~1.75E-2~&~7.36E-2~&~$2.30$~&~1.45~&~1.09~\\
    ~122~&~$0.64$~&~$512^3$~&~4.20E-3~&~0.74~&~1.78E-2~&~7.55E-2~&~$2.29$~&~1.46~&~1.10~\\
    ~122~&~$0.77$~&~$512^3$~&~4.23E-3~&~0.69~&~1.82E-2~&~7.84E-2~&~$2.27$~&~1.48~&~1.13~\\
    ~119~&~$0.89$~&~$512^3$~&~4.27E-3~&~0.70~&~1.83E-2~&~7.84E-2~&~$2.27$~&~1.48~&~1.13~\\
    ~117~&~$1.00$~&~$512^3$~&~4.31E-3~&~0.63~&~1.89E-2~&~8.28E-2~&~$2.23$~&~1.51~&~1.17~\\
  \end{tabular}
  \caption{Simulation parameters:
   Taylor Reynolds number $R_{\lambda}$,
   turbulent Mach number $M_t$,
   grid points $N^3$,
   kinematic viscosity $\nu$,
   average dissipation rate of kinetic turbulent energy $\varepsilon$,
   Kolmogorov length scale $\eta$,
   Kolmogorov time scale $\tau_{\eta}$,
   root-mean-square fluctuation velocity $u_{rms}=\sqrt{\langle u_iu_i \rangle}$,
   integral length scale $L_f=3\pi /(2u_{rms}^2) \int_0^{+\infty} [E(k)/k]dk$
   and large-scale eddy turnover time $T_f=\sqrt{3}L_f/u_{rms}$.}
  \label{tab:param_f}
  \end{center}
\end{table}

The light particles are regarded as point objects,
neglecting the influence of particles on turbulent flows
and their mutual interaction.
The motion of light particles is governed by:
\begin{equation}
 \frac{\mathrm{d} \textbf{\emph{x}}_p}{\mathrm{d} t}=\textbf{v}_p
 \label{eq:dx_dt}
\end{equation}
\begin{equation}
 \frac{\mathrm{d}\textbf{v}_p}{\mathrm{d}t}=\frac{\textbf{\emph{u}}_p-\textbf{v}_p}{\tau_p}
 +\beta \frac{\mathrm{D}\textbf{\emph{u}}_p}{\mathrm{D}t}
 \label{eq:dv_dt}
\end{equation}
\begin{equation}
 \tau_p=\frac{\rho_f d^2}{12\beta\mu}
 \label{eq:tau_p}
\end{equation}
\begin{equation}
 \beta=\frac{3\rho_f}{2\rho_p+\rho_f}
 \label{eq:beta}
\end{equation}
where $\textbf{\emph{x}}_p$, $\textbf{v}_p$,
$\tau_p$ and $d$ are the position, velocity,
response time and diameter of particles, respectively,
and $\textbf{\emph{u}}_p$ is the velocity of the flow
at the position of the particle.
$\mu$ is the dynamic viscosity of fluid
at the position of particles, and we notice that
both $\tau$ and $\beta$ are dependent on space and time
due to the variation of the underlying fluid velocity $\rho_f$.
As a result, in a compressible flow,
even particle with a density $\rho_p$ equal to the average flow density
$\langle \rho_f \rangle$ will nevertheless be locally subjected to
inertial forces due to the variation of the local flow density.
In our case we have chosen $\rho_p = 0.01 \langle \rho_f \rangle$
for the light particles. It is important to stress
that Eq. (\ref{eq:dv_dt}) must be considered an approximation
of the true point-like description of inertial forces.
Beside corrections present also in incompressible forces,
as the Faxen and Basset terms \citep{maxey1983equation,
gatignol1983faxen}, here we need to consider also new forces induced by
density variations of the fluid flow at the particle positions
\citep{parmar2011generalized,parmar2012equation}.
In appendix A we present a thoughtful discussion
about these new effects and show that at moderate Mach numbers,
as the one here investigated,
they can be most of the time safely neglected.
The fluid velocity at the position of particles,
$\textbf{\emph{u}}_p$, is evaluated
by a fourth-order Lagrangian interpolation.
The operator, $\mathrm{D}\slash \mathrm{D}t$,
represents the material derivative.
$256^3$ particles are uniformly released
into the computational domain
after the flow reaches a statistically steady state.
After the particles are fully mixed,
{we sample over 15 times large eddy turnover time ($T_f$)
to explore the spatial, instantaneous statistics of particles.}
The Stokes number ($\mathrm{St}=\tau_p/\tau_{\eta}$)
of particles is approximately 0.21,
and the average of $\beta$ is around 2.94.
{In addition, tracer particles whose velocity  follows
the local fluid velocity are also analyzed for comparison.}

\section{Results and analysis}\label{sec:resultsandanalysis}
\begin{figure}
 \centering
  \includegraphics[width=0.48\textwidth]{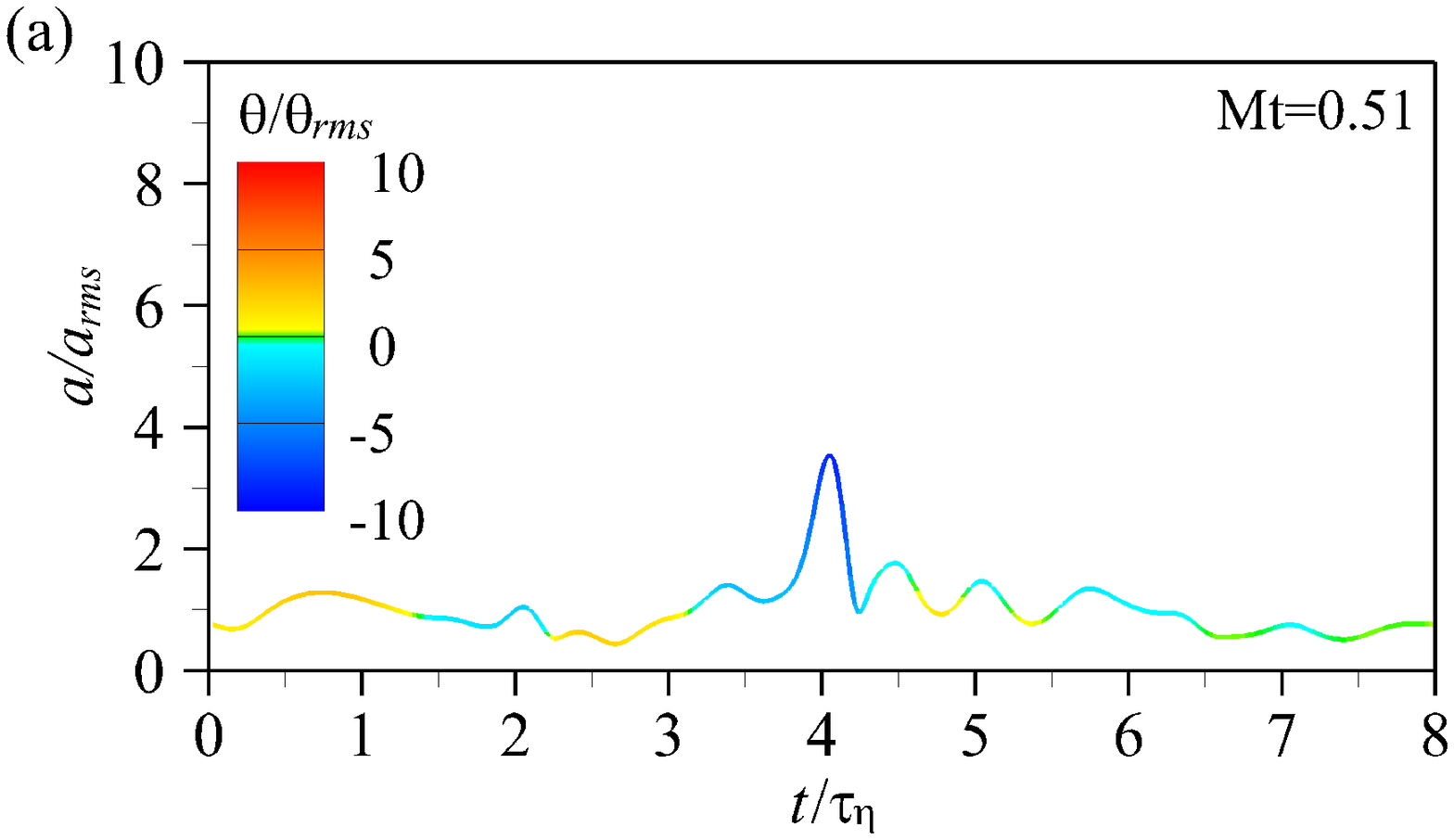}
  \includegraphics[width=0.48\textwidth]{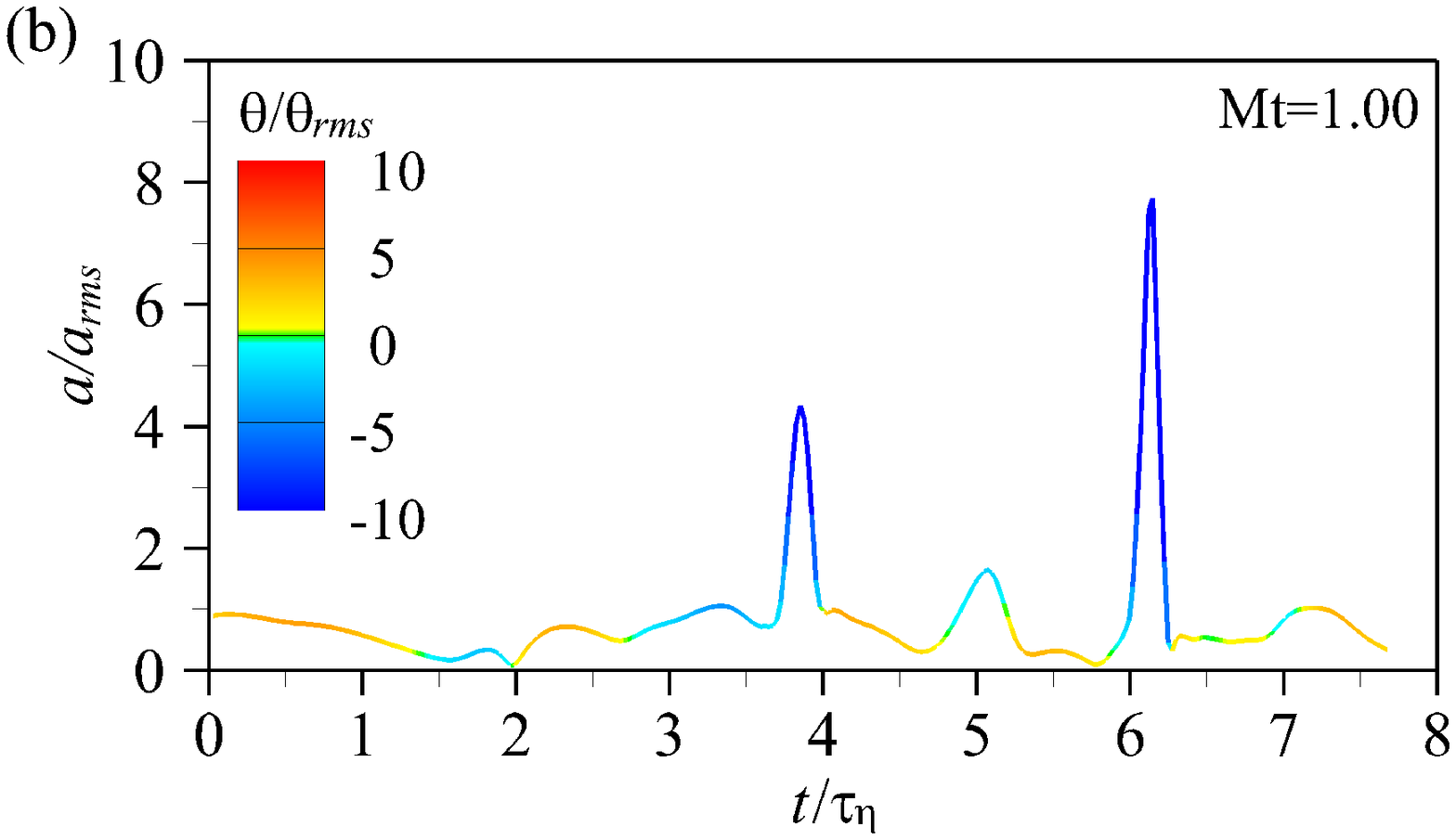}

  \includegraphics[width=0.48\textwidth]{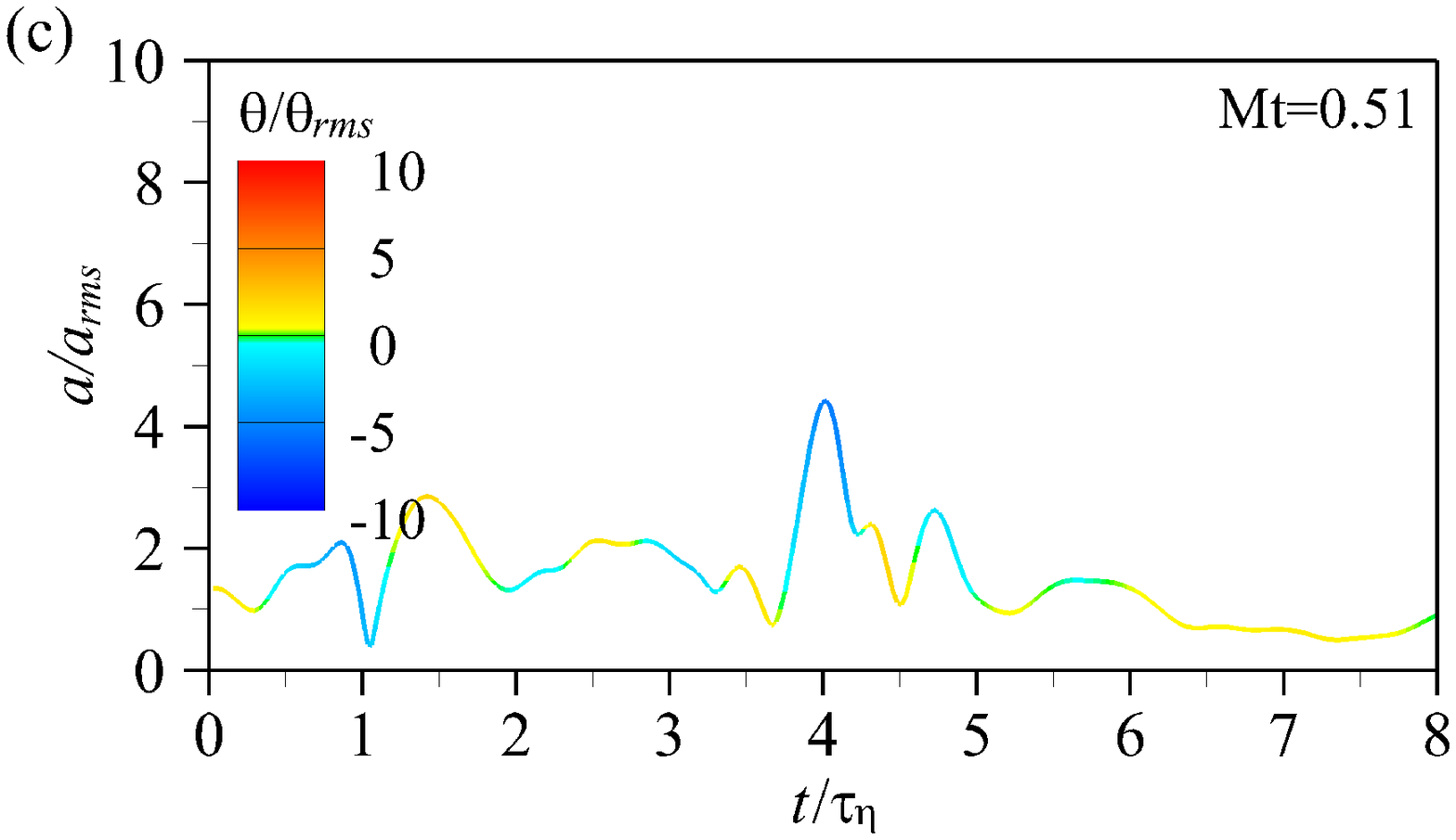}
  \includegraphics[width=0.48\textwidth]{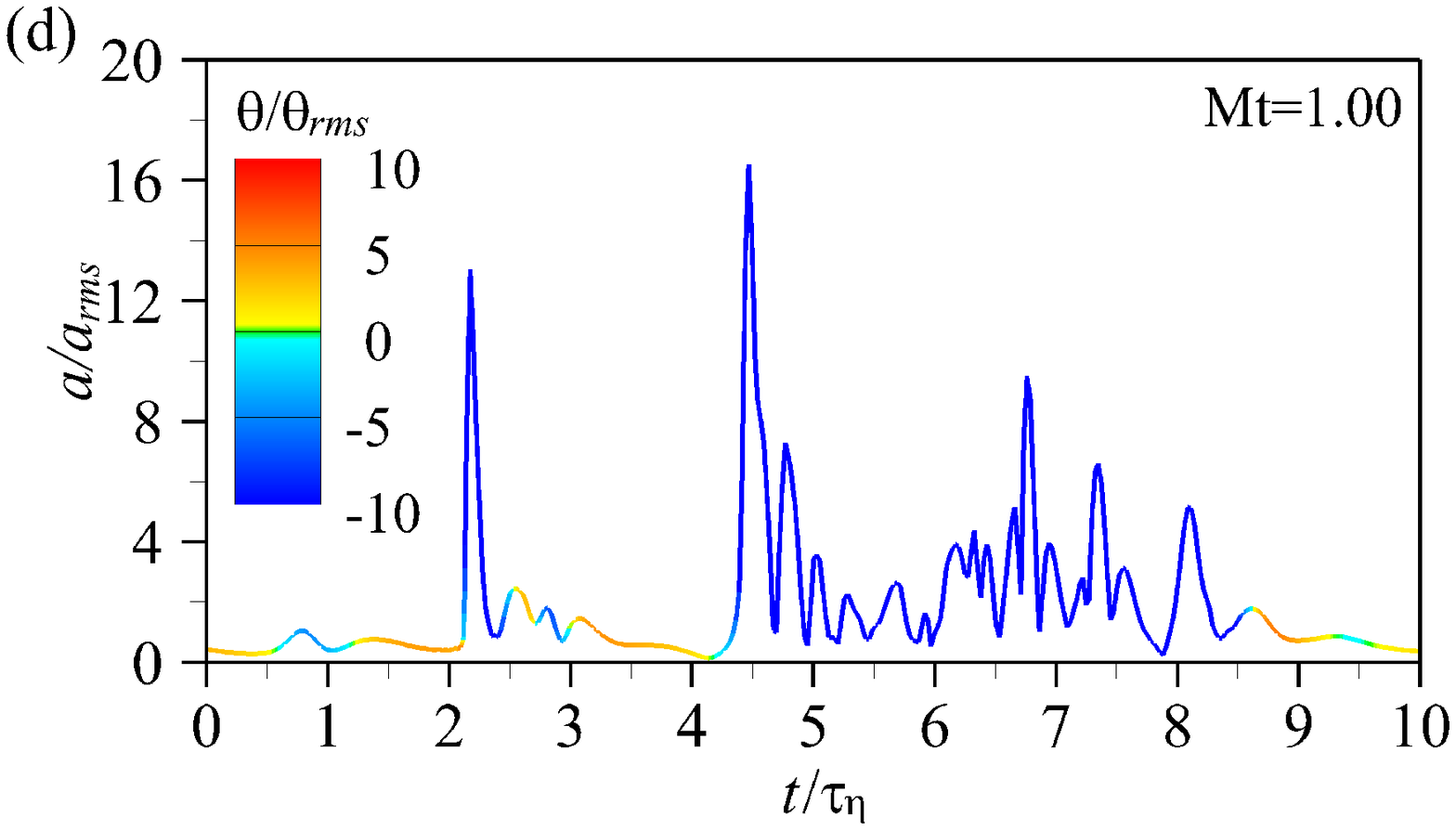}
  \caption{The evolution of the acceleration magnitude, $a$,
  of typical tracers  (a-b) and light particles (c-d) at $M_t=0.51$ and $M_t=1.00$.
  The colorbar indicates the  divergence of fluid velocity, $\theta$,
  at the position of particles. Here $a_{rms}$ and $\theta_{rms}$ separately represent
  the root mean square value of $a$ and $\theta$.}
\label{fig:accp_time_theta}
\end{figure}
Before presenting the results from the statistical analysis,
we show  in Figure \ref{fig:accp_time_theta} some characteristic
evolution of the  acceleration magnitudes, $a$,
along tracers and light particles trajectories
near strong compression structures (e.g. shocklets)
at $M_t=0.51$ and $M_t=1.00$, color coded in terms of
the local fluid compressibility $\theta$.
For tracer particles, there is no significant difference between $M_t=0.51$ and 1.00,
except for the larger peak of acceleration magnitudes near shocklets at $M_t=1.00$.
For light particles at $M_t=0.51$, the evolution of acceleration magnitudes
is similar to that of tracer particles.
However, at $M_t=1.00$, light particles seem to be trapped
near the shocklets for a long time (several times $\tau_{\eta}$),
and the acceleration magnitudes of particles fluctuate dramatically.
\begin{figure}
\centering
\centering
  \includegraphics[width=0.48\textwidth]{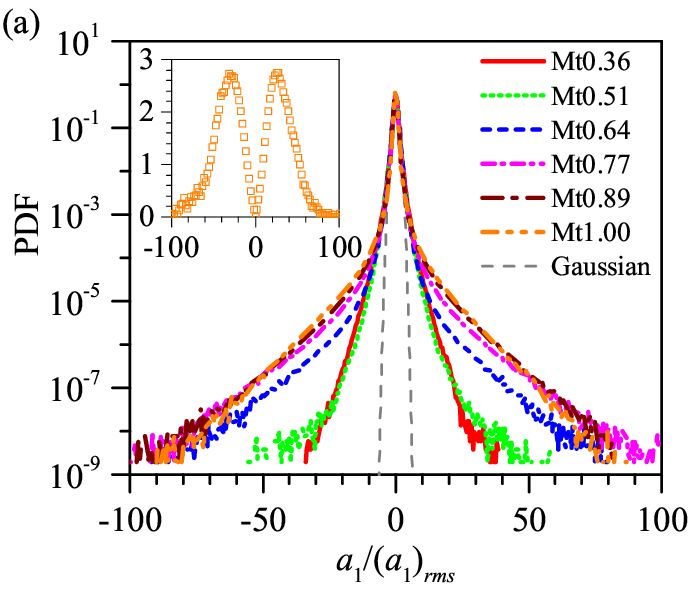}
  \includegraphics[width=0.48\textwidth]{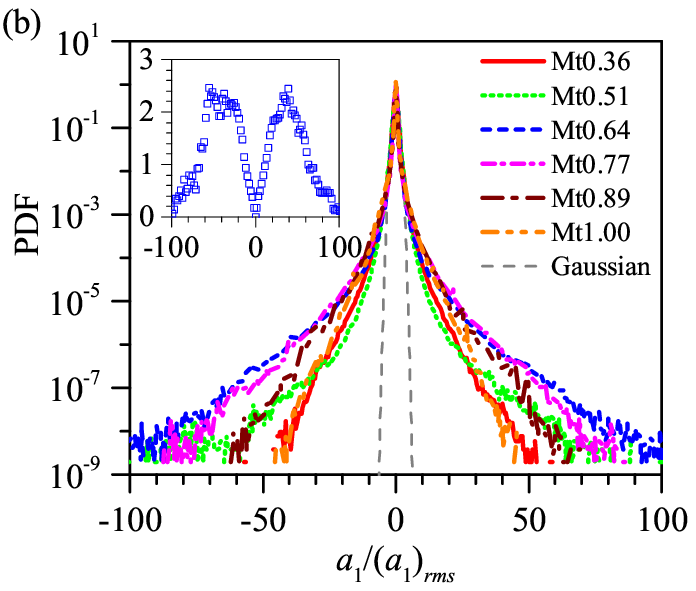}
  \caption{The probability density functions (PDFs), $P(a_{i})$, of one component of the accelerations $a_i$, normalized with its root mean square. Because of isotropy all three components are equivalent. For clarity here we show only the case with  $i=1$.
   (a) is for tracer particles. Inset:  $(a_1)^4P(a_1)$, at $M_t=1.00$, is presented to check the statistical convergence of the fourth-order moments; (b) is for light particles. Inset: $(a_1)^4P(a_1)$ is shown at $M_t=0.64$.}
\label{fig:Apx_PDF}
\end{figure}

\begin{figure}
  \centering
  \includegraphics[width=0.48\textwidth]{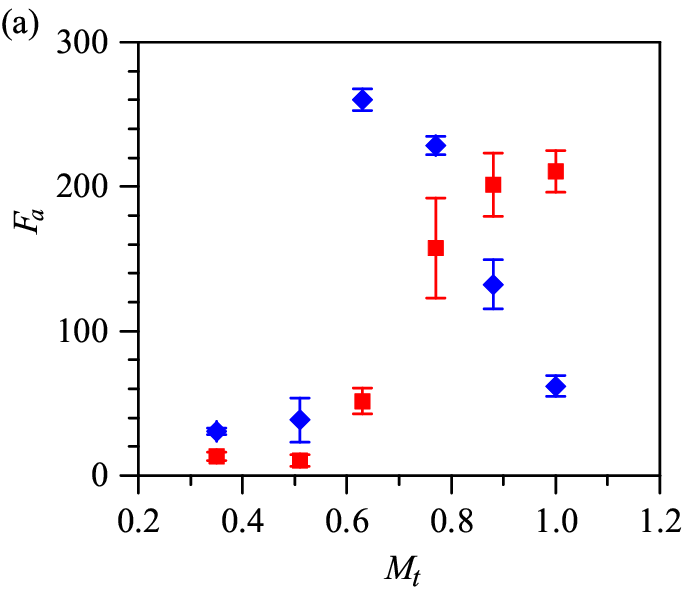}
  \includegraphics[width=0.48\textwidth]{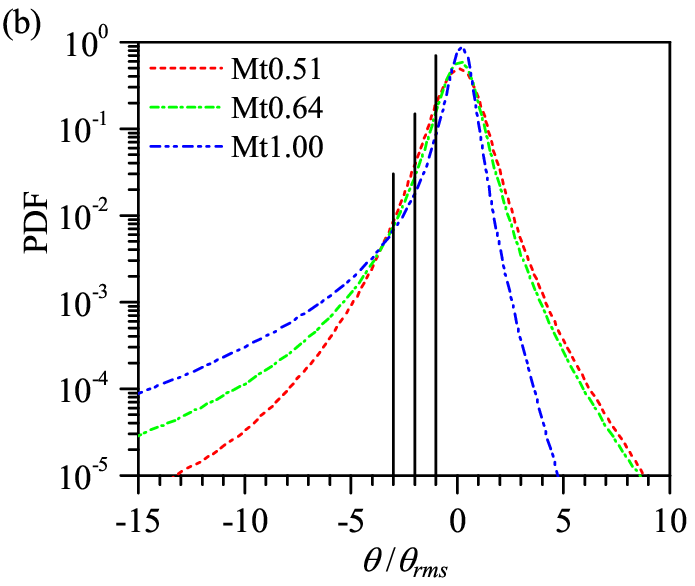}
  \caption{Panel (a): the flatness factor of particle acceleration components, $F_a$,
  at changing $M_t$ for tracers (red squares) and light particles (blue diamonds).
  The error bar is calculated from the scatter among 30 subsamples obtained from 30 time snapshots.
  Panel (b): the PDFs of  compressibility $\theta$ in the whole computational domain
  at $M_t=0.51$, $0.64$ and $1.00$. The three vertical lines mark the  values CR1, CR2 and CR3
  where  $\theta < -\theta_{rms}$, $\theta < -2\theta_{rms}$ and $\theta < -3\theta_{rms}$.}
  \label{fig:flatness_theta_PDF}
\end{figure}

\begin{figure}
\centering
\centering
  \includegraphics[width=0.32\textwidth]{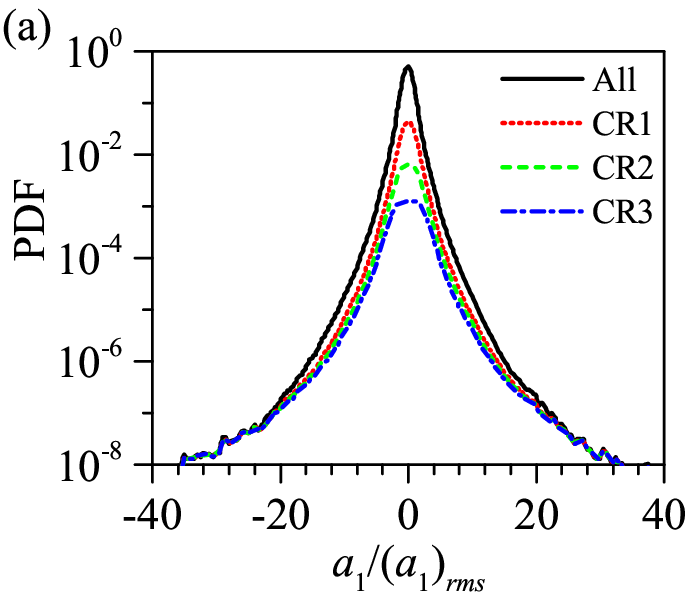}
  \includegraphics[width=0.32\textwidth]{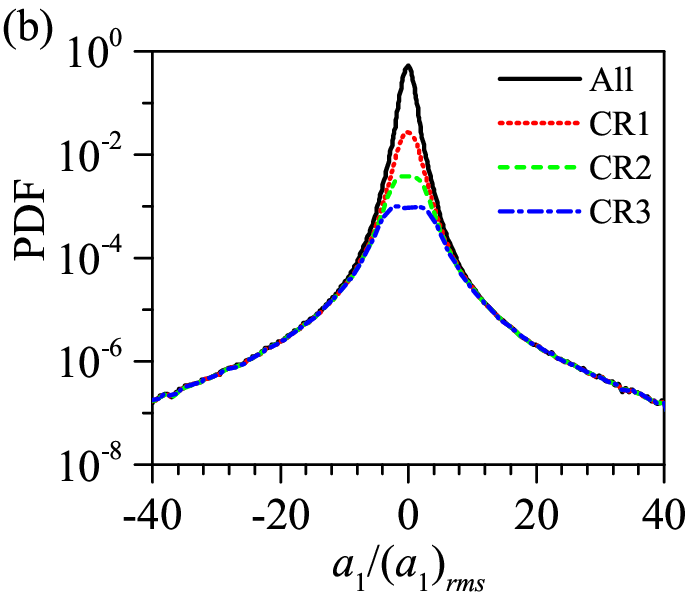}
  \includegraphics[width=0.32\textwidth]{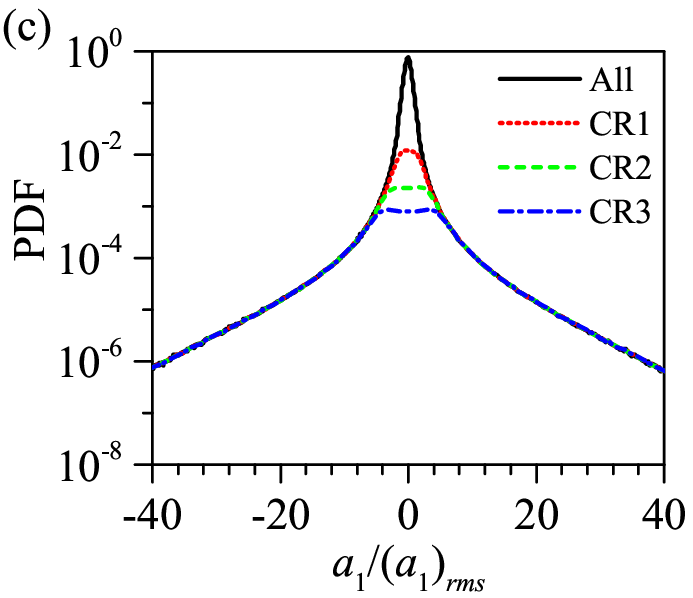}

  \includegraphics[width=0.32\textwidth]{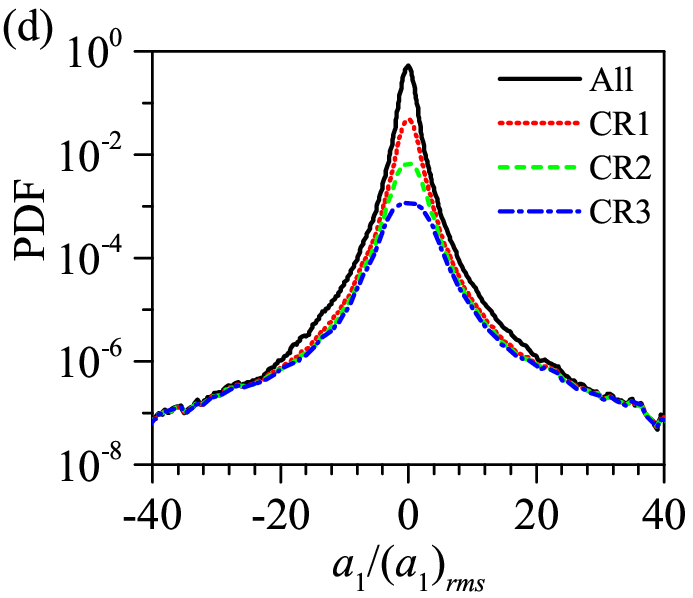}
  \includegraphics[width=0.32\textwidth]{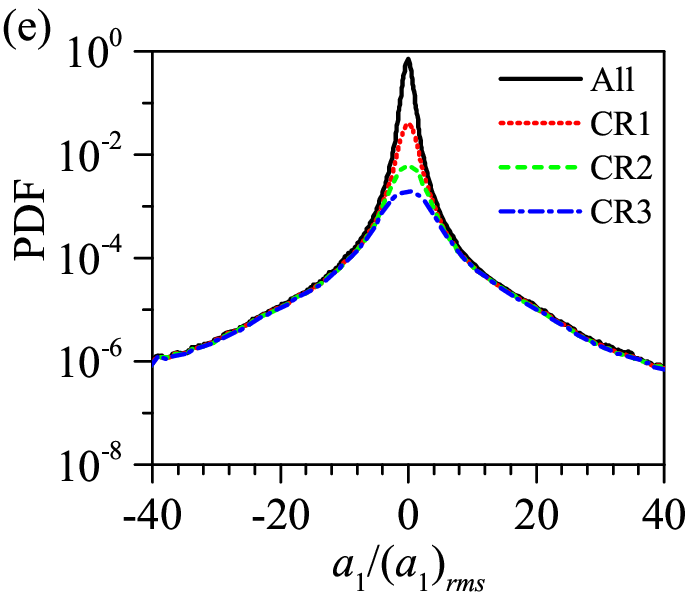}
  \includegraphics[width=0.32\textwidth]{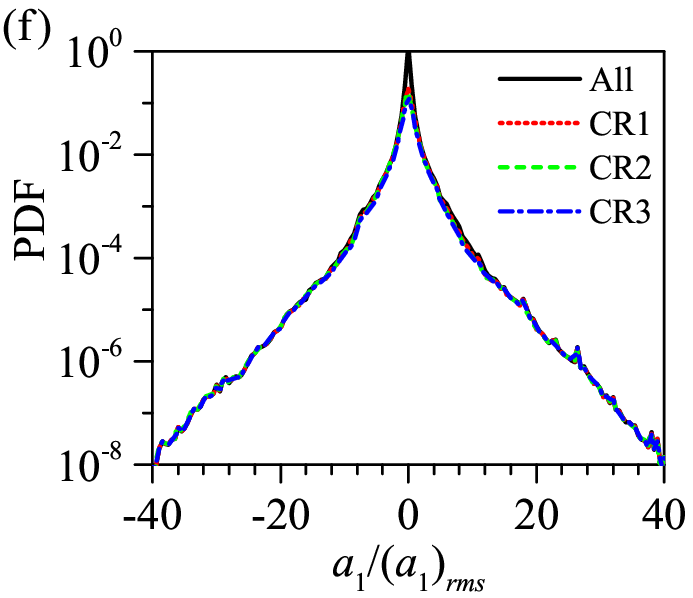}
  \caption{ The probability density functions (PDFs) of particle acceleration components.
  (a)-(c) is for tracer particles at $M_t$=0.51, 0.64 and 1.00 respectively;
  (d)-(f) is for light particles at $M_t$=0.51, 0.64 and 1.00.
  The black solid line is the PDF of $a_1$ for particles in the whole computational domain;
  the red dotted line, green dashed line and blue dash-dotted line represent
  the contribution of particles in CR1, CR2 and CR3 separately.}
\label{fig:Apx_PDF_theta}
\end{figure}

\begin{figure}
\centering
  \includegraphics[width=0.48\textwidth]{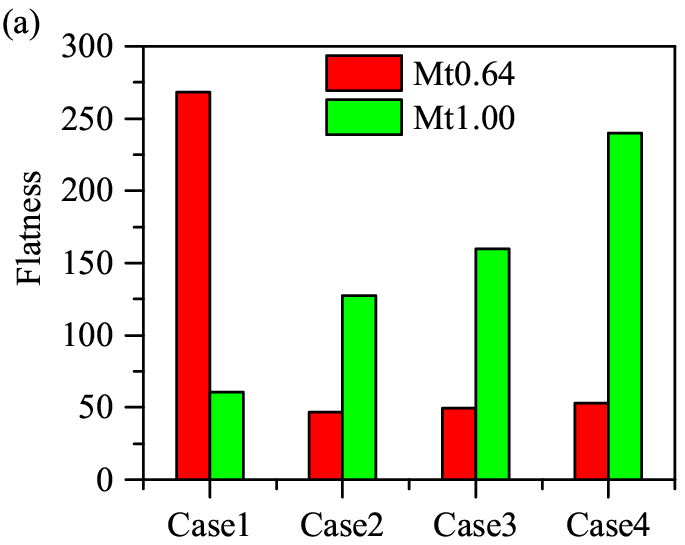}
  \includegraphics[width=0.48\textwidth]{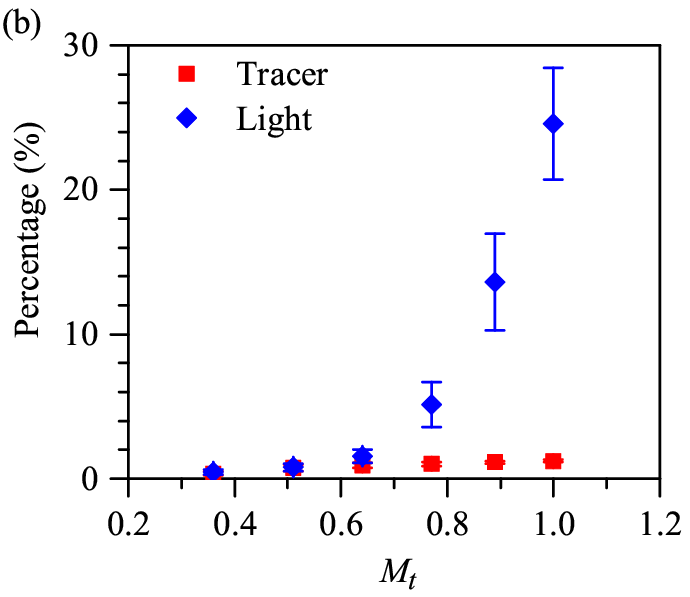}
  \caption{Panel (a): $F_a$ of light particles in different regions.
  Here, Case 1: the whole domain;
  Case 2: the regions with $\theta > -3\theta_{rms}$;
  Case 3: the regions with $\theta > -2\theta_{rms}$;
  Case 4: the regions with $\theta > -\theta_{rms}$.
  Panel (b): the ratio of the particle number in CR3 to the total particle number in the whole domain.}
  \label{fig:filter_percentage}
\end{figure}

The probability density functions (PDFs) of
one component of the particle accelerations,
$a_i$, with $i=1, 2, 3$, are illustrated
in Fig. \ref{fig:Apx_PDF}.
It is apparent that all PDFs have some
intense stretched tails, indicating that
the accelerations of tracer and light
particles are both intermittent.
The strength of intermittency can be
quantified by the flatness (or kurtosis).
For tracer particles, the flatness of acceleration components,
$F_a$, steadily increases with the increase of $M_t$,
as seen in Fig. \ref{fig:flatness_theta_PDF} (a).
Here, $F_a$ is defined as:
$F_a = \langle (a_1)^4 \rangle \slash \langle  (a_1)^2 \rangle^2$.
The operator $\langle \cdot \rangle$ means the average among particles.
However, $F_a$ of light particles does not follow this trend.
As $M_t$ increases from 0.36 to 1.00,
$F_a$ of light particles firstly increases,
then followed by a reduction after $M_t$ exceeds about 0.6.
{In order to explain this phenomenon,
we define the compression regions: CR1, CR2 and CR3
representing the regions with  $\theta < -\theta_{rms}$,
$\theta < -2\theta_{rms}$ and $\theta < -3\theta_{rms}$ separately,
as shown in Fig. \ref{fig:flatness_theta_PDF} (b).}
Then, it is found that the stretched tails
of acceleration PDFs predominantly result from the particles
in strong compression regions when $M_t$ is large,
as shown in Fig. \ref{fig:Apx_PDF_theta}.
The increase of $F_a$ of tracer particles
(also applicable for light particles at $M_t < 0.6$),
is attributed to the fact that increasing compressibility
increases the flow intermittency too.
However, when $M_t$ is larger than 0.6, why does $F_a$
of light particles decrease with the increase of $M_t$?
In the following we argue that it is related to certain phenomena
of light particles in compression regions.

\begin{figure}
\centering
\centering
  \includegraphics[width=0.45\textwidth]{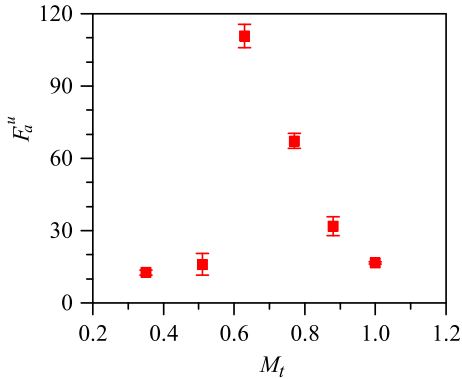}
  \caption{The flatness factor of acceleration components of fluid elements
  at the position of light particles, $F_a^u$, versus $M_t$.}
\label{fig:flat_validation}
\end{figure}
Let us first compare the $F_a$ of light particles in different regions,
as shown in Fig. \ref{fig:filter_percentage} (a),
where Case 1 represents the whole domain;
Case 2 represents the region with $\theta>-3\theta_{rms}$
(removing CR3 from Case 1);
Case 3 represents the region with $\theta>-2\theta_{rms}$
(removing CR2 from Case 1);
Case 4 represents the region with $\theta>-\theta_{rms}$
(removing CR1 from Case 1).
We can notice that at $M_t=0.64$, $F_a$ decreases sharply
once the light particles in CR3 are removed.
In contrast, at $M_t=1.00$, $F_a$ increases steadily
as the particles in CR3, CR2 and CR1 are removed.
To explain this difference,
we further measure the number of light particles
in CR3, see Fig. \ref{fig:filter_percentage} (b),
where the number of tracer particles in CR3 slightly increases
as $M_t$ increases from 0.36 to 1.00.
However, the number of light particles in CR3
dramatically increases when $M_t>0.6$.
Especially at $M_t=1.00$, it accounts for approximate $25\%$
of the total particle number in the whole domain.
According to Fig. \ref{fig:Apx_PDF_theta}, ones can understand
that the extremely large accelerations of particles
mainly come from the strong compression regions (such as CR3).
Then, if the number of particles in CR3 is small (like tracers),
the probability of the large accelerations is still small.
Therefore, a slight increase of particle number in CR3 will make a
contribution to the enhancement of the acceleration intermittency.
Hence, $F_a$ of tracer particles increases with the increase of $M_t$.
On the other hand, if the number of particles in CR3 is too large
(like light particles at $M_t = 0.77-1.00$),
there will be a large portion of particles with large accelerations.
Namely, most of them cannot be regarded as rare events any more.
Then, the increase of particle number in CR3 will not contribute to
the increase of acceleration intermittency any more.
Mathematically, large amount of particles accumulating in CR3 will cause
a significant increase of the standard deviation of accelerations ($\sigma_a$).
And the increase of $(\sigma_a)^4$ is faster than the increase of
the forth-order moment of accelerations.
In conclusion, the preferential concentration of light particles
in strong compression regions (e.g. CR3) results in the decrease
of $F_a$ when $M_t > 0.6$.
In order to further validate this mechanism, the flatness factor of
acceleration components of fluid elements at the position of
light particles, $F_a^u$, is measured, shown in Fig. \ref{fig:flat_validation}
where $F_a^u$ varies with the same trend as $F_a$ of light particles.

\subsection{Longitudinal acceleration at different $M_t$}\label{subsec:Ap_L}
\begin{figure}
\centering
\centering
  \includegraphics[width=0.45\textwidth]{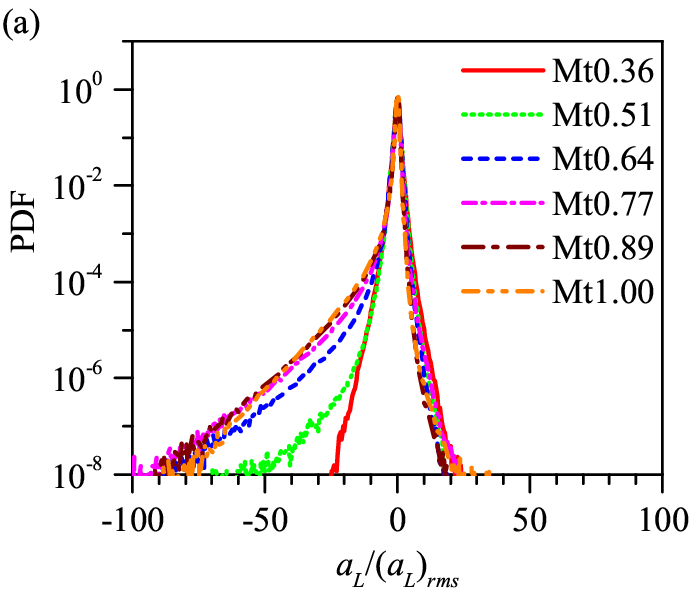}
  \includegraphics[width=0.45\textwidth]{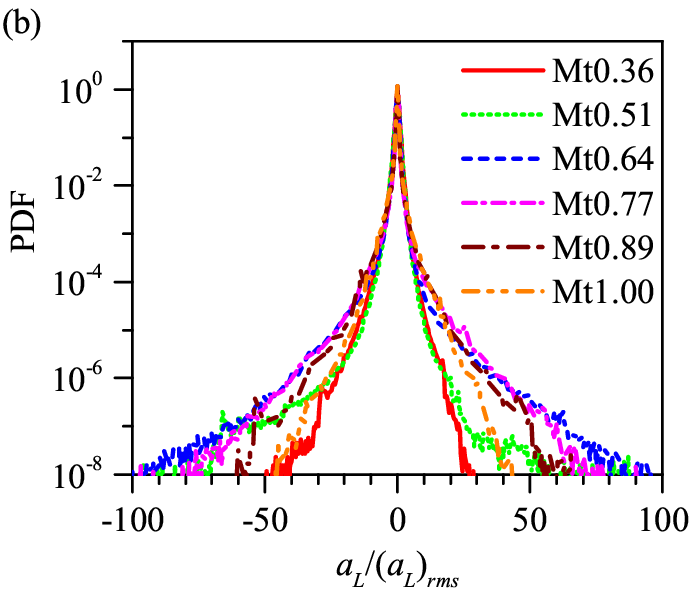}
  \caption{The PDFs of longitudinal accelerations.
  (a) is for tracer particles; (b) is for light particles.}
\label{fig:Ap_L_PDF}
\end{figure}

\begin{figure}
\centering
\centering
  \includegraphics[width=0.45\textwidth]{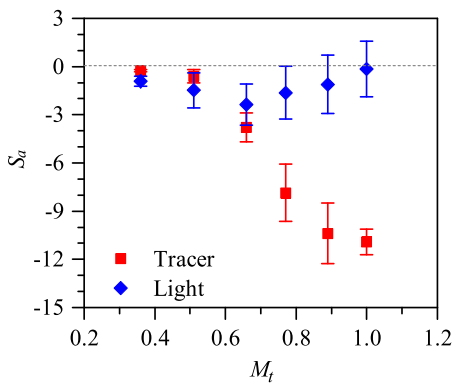}
  \caption{The skewness factor of particle longitudinal accelerations,
  $S_a$, varies with $M_t$.}
\label{fig:Ap_L_skew}
\end{figure}

Figure \ref{fig:Ap_L_PDF} presents
the PDFs of particle longitudinal accelerations
($\textbf{\emph{a}}_L$) at $M_t = 0.36-1.00$.
Ones can see that for tracer particles,
the PDFs of $\textbf{\emph{a}}_L$ all skew to the negative value.
However, for light particles,
the PDFs of $\textbf{\emph{a}}_L$ only moderately skew to
the negative value at low $M_t$ (e.g. $M_t=0.51$),
and this trend vanishes as the PDFs become almost symmetric
at high $M_t$ (e.g. $M_t=1.00$).
To quantify this difference, we measure the skewness factor of
$\textbf{\emph{a}}_L$, $S_a$, as shown in Fig. \ref{fig:Ap_L_skew}.
For tracer particles, $S_a$ becomes more negative
as $M_t$ increases from 0.36 to 1.0.
As well known, shocklets will appear
in compression regions with the increase of $M_t$.
When tracer particles pass through the shocklets
from upstream to downstream, they will be decelerated significantly
owing to the negative pressure gradient \citep{anderson2010fundamentals}.
As $M_t$ increases, the shocklets become stronger
(the deceleration is also more remarkable).
Therefore, the PDF of $\textbf{\emph{a}}_L$
is more skewed to negative value
and $S_a$ becomes more negative for tracer particles
when $M_t$ increases.

\begin{figure}
\centering
\centering
  \includegraphics[width=0.45\textwidth]{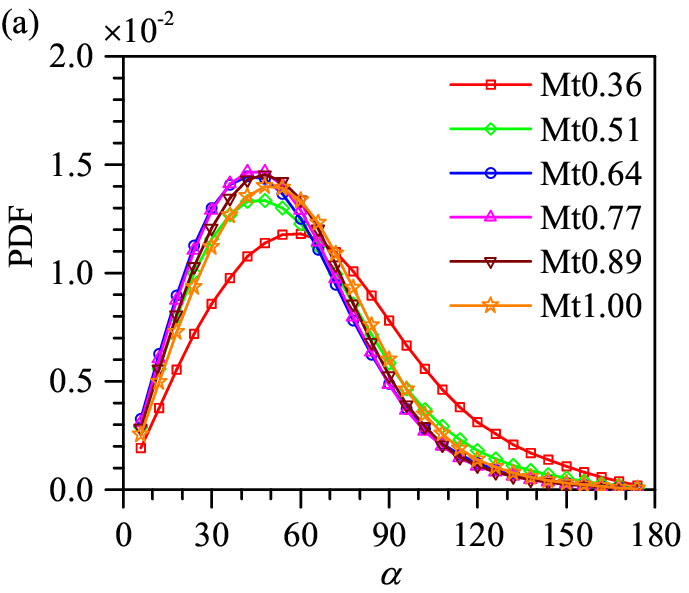}
  \includegraphics[width=0.45\textwidth]{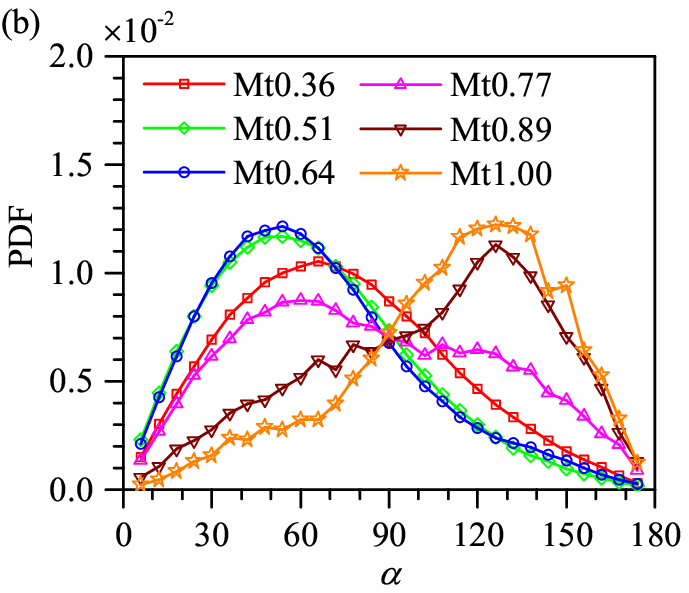}
  \caption{The PDFs of angle ($\alpha$) between the velocity of particles
  and the pressure gradient at the position of particles in compression regions.
  (a) is for tracer particles; (b) is for light particles.}
\label{fig:Angle}
\end{figure}

\begin{figure}
\centering
\centering
  \includegraphics[width=0.45\textwidth]{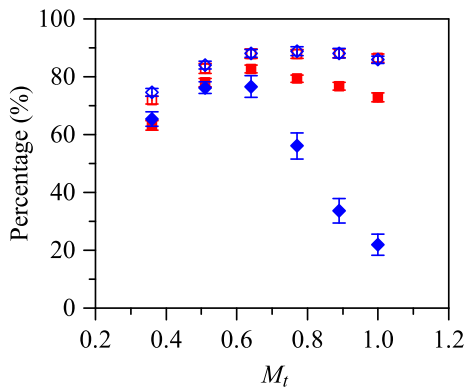}
  \caption{The ratio of $N_{CR1}^{in}$ (red squares) and $N_{CR1}$ (blue diamonds) to the total number of particles in CR1.
  The empty symbols are for tracer particles and the filled symbols
  are for light particles.}
\label{fig:proportion}
\end{figure}

As shown in Fig. \ref{fig:Ap_L_skew}, for light particles,
$S_a$ also decreases with $M_t$ when $M_t$ is small,
similar to that of tracer particles.
However $S_a$ gradually goes back to zero after $M_t > 0.6$.
A similar phenomenon was suggested by \citet{yang2014interactions},
who discovered that $S_a$ of light particles in compression regions
is close to 0 at $M_t \approx 1.03$,
and there are many light particles whose velocities have an obtuse angle
with local pressure gradient.
To understand the above difference between tracers and light particles,
we show in Fig. \ref{fig:Angle} the PDF of the angle
between particle velocity and pressure gradient, $\alpha$, in CR1
(compression region with $\theta<-\theta_{rms}$).
Ones can see that most tracer particles in CR1
have an angle of $\alpha < 90^{\circ}$ at all $M_t$.
However, for light particles, the number of particle with $\alpha > 90^{\circ}$
steadily increases as $M_t$ increases from 0.6 to 1.0.
We then track the particles in CR1
and statistically measure how many of them have an angle of $\alpha < 90^{\circ}$
when they enter CR1, denoted as $N_{CR1}^{in}$.
Also, the number of particles with $\alpha < 90^{\circ}$
in CR1 at one time point is statistically measured,
defined as $N_{CR1}$.
Figure \ref{fig:proportion} shows that for tracer particles,
$N_{CR1}$ is always close to $N_{CR1}^{in}$ at different $M_t$.
Namely, tracer particles rarely change their direction of movement in CR1.
However, for light particles,
$N_{CR1}$ is noticeably smaller than $N_{CR1}^{in}$ at high $M_t$,
and this deviation becomes more significant as $M_t$ increases.
That is, as $M_t$ increases, more light particles reverse
their direction of velocity from $\alpha < 90^{\circ}$
to $\alpha > 90^{\circ}$ in CR1, so that more light particles
which were decelerated turn to be accelerated.
This is the main reason for the PDF of $\textbf{\emph{a}}_L$
being more symmetric when $M_t>0.6$.
Besides, $N_{CR1}$ of light particles decreases moderately
as $M_t$ increases from 0.6 to 1.0, which implies that
at high $M_t$, more light particles have positive
longitudinal accelerations when they are entering CR1.
It is also a factor that $S_a$ of light particles goes back to 0.

Why do many light particles reverse their velocity direction
from $\alpha < 90^{\circ}$ to $\alpha > 90^{\circ}$ in CR1,
but few tracer particles do?
It can be seen in Eq. (\ref{eq:dv_dt}) that the effect
of local fluid acceleration(the second term
on the right-hand side) on light particles was amplified by $\beta$
(for light particles, $\beta \approx 2.94$ in our simulations).
Thus, light particles are easier to be decelerated by
the negative pressure gradient near shocklets
as the dominant contribution to fluid acceleration comes from pressure gradient.
After light particles enter compression regions from upstream,
their direction of velocity is more likely be reversed near skocklets,
and this trend will be more remarkable as $M_t$ increases.

\subsection{Transverse acceleration at different $M_t$}\label{subsec:Ap_T}
\begin{figure}
\centering
\centering
  \includegraphics[width=0.45\textwidth]{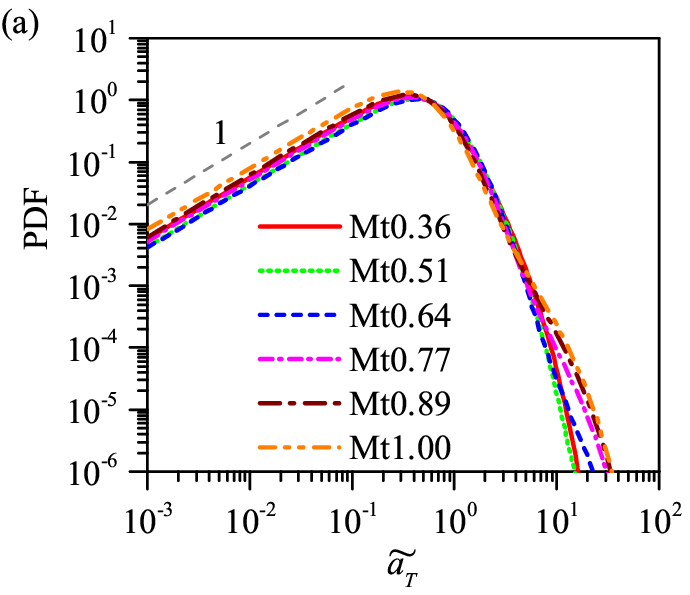}
  \includegraphics[width=0.45\textwidth]{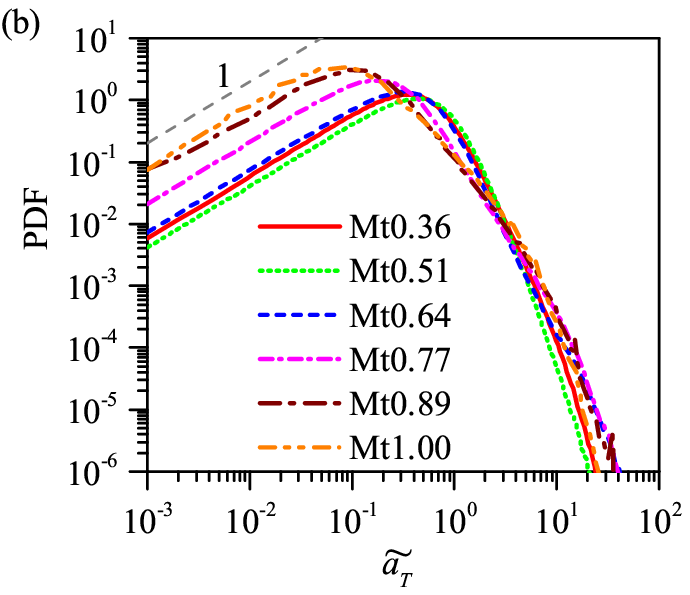}
  \caption{The PDFs of $\widetilde{a}_T$
  at different $M_t$. (a) is for tracer particles; (b) is for light particles.}
\label{fig:Ap_T_PDF}
\end{figure}

The change of velocity direction is intimately linked to
the transverse acceleration ($\textbf{\emph{a}}_T$).
In Figure \ref{fig:Ap_T_PDF} (a) we show
the PDFs of normalized transverse acceleration magnitudes,
$\widetilde{a}_T=a_T/ (a_T)_{rms}$, of tracer particles at different $M_t$,
where $a_T$ is the magnitude of transverse acceleration.
When $\widetilde{a}_T$ is not large (e.g. $\widetilde{a}_T<3$),
the PDFs of $\widetilde{a}_T$ are close to each other
for the cases with different $M_t$.
However, when $\widetilde{a}_T>3$,
the PDFs display slightly wider tails at higher $M_t$.
This implies that when $\widetilde{a}_T$ is small,
it is mainly influenced by vortices, rather than shocklets.
The exist of shocklets only increases the probability
of large $\widetilde{a}_T$.
On the contrast, the PDFs of $\widetilde{a}_T$
of light particles are quite distinguishing at different $M_t$,
as illustrated in Fig. \ref{fig:Ap_T_PDF} (b)
where at higher $M_t$, the probability of small $\widetilde{a}_T$ is larger,
indicating that  $\widetilde{a}_T$ of light particles
is easier to be influenced by shocklets than that of tracer particles.
In addition, when $\widetilde{a}_T$ is quite small,
the PDFs all linearly scale with $\widetilde{a}_T$,
which can be explained by the model developd by  \citet{xu2007curvature}.
The three acceleration components of particles can be regarded as
independent Gaussian random variables
when their magnitude is quite close to zero.
Then, $a_T$ of particles will follow a $\chi$ distribution
with 2 degrees of freedom, leading to a linear dependency
for small enough  $\widetilde{a}_T$.

\begin{figure}
\centering
\centering
  \includegraphics[width=0.45\textwidth]{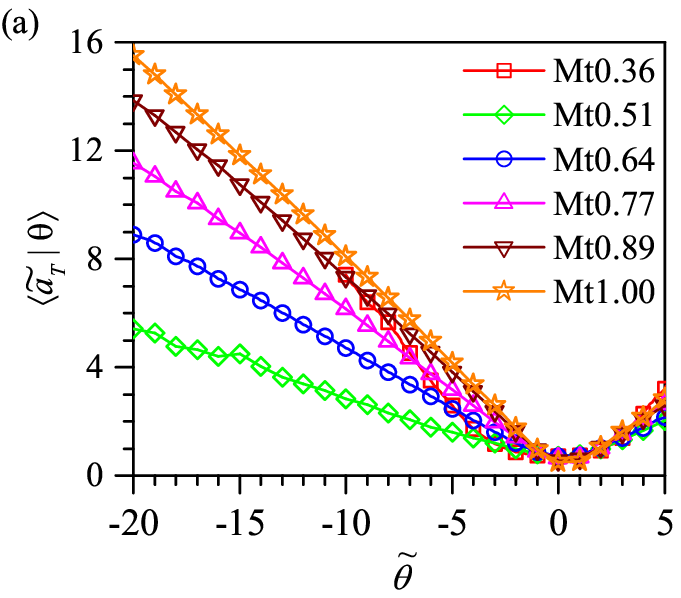}
  \includegraphics[width=0.45\textwidth]{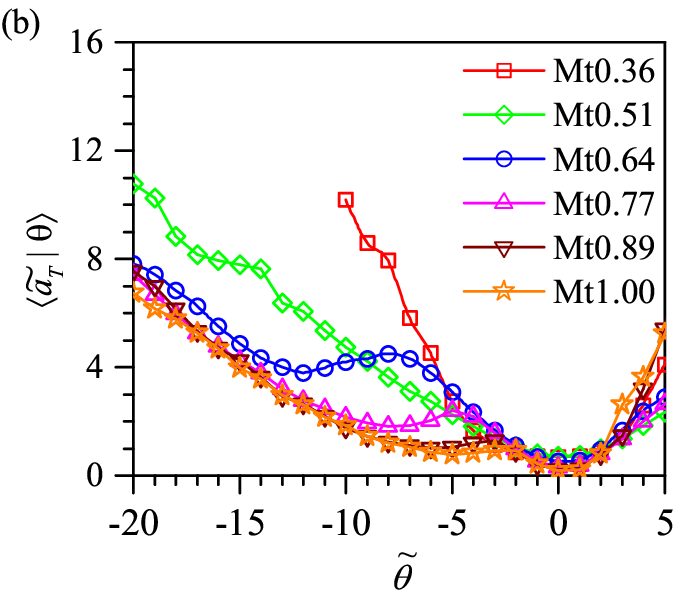}
  \caption{$\langle \widetilde{a}_T | \theta \rangle$ varies with
  $\tilde{\theta}$ at different $M_t$. (a) is for tracer particles;
  (b) is for light particles.}
\label{fig:Ap_T_theta}
\end{figure}
Furthermore, we have measured the average
of $\widetilde{a}_T$ conditioned for a given velocity divergence,
$\langle \widetilde{a}_T | \theta \rangle$,
shown in Fig. \ref{fig:Ap_T_theta}.
For tracer particles, except for $M_t=0.36$,
$\langle \widetilde{a}_T | \theta \rangle$
linearly increases with the decrease of $\tilde{\theta}$
($\tilde{\theta}=\theta / \theta_{rms}$)
in compression regions, and it gradually increases
with the increase of $M_t$, for a given $\tilde{\theta}$.
On the other hand, for light particles,
the linear increase of $\langle \widetilde{a}_T | \theta \rangle$
with the decrease of $\tilde{\theta}$ does not appear in compression regions.
Particularly when $M_t > 0.6$, there is always a section
where $\langle \widetilde{a}_T | \theta \rangle$ decreases as $\tilde{\theta}$ decreases.

\subsection{Autocorrelation function of acceleration components at different $M_t$}\label{subsec:Autocorltn}
\begin{figure}
\centering
\centering
  \includegraphics[width=0.49\textwidth]{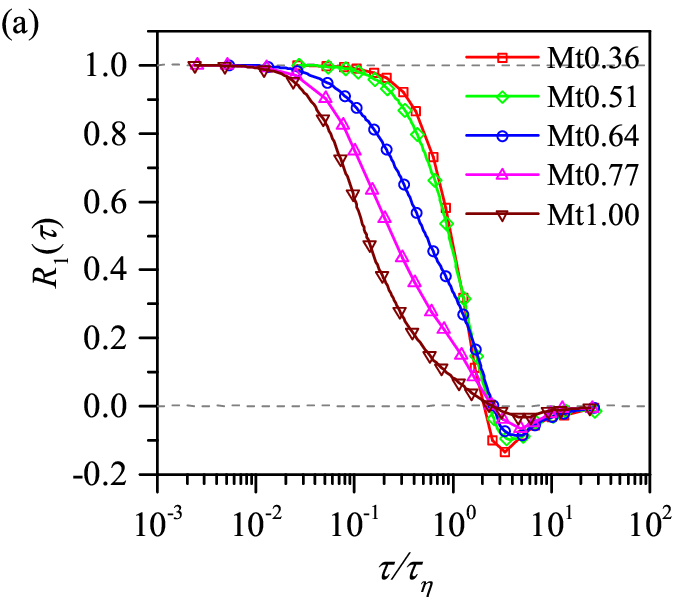}
  \includegraphics[width=0.49\textwidth]{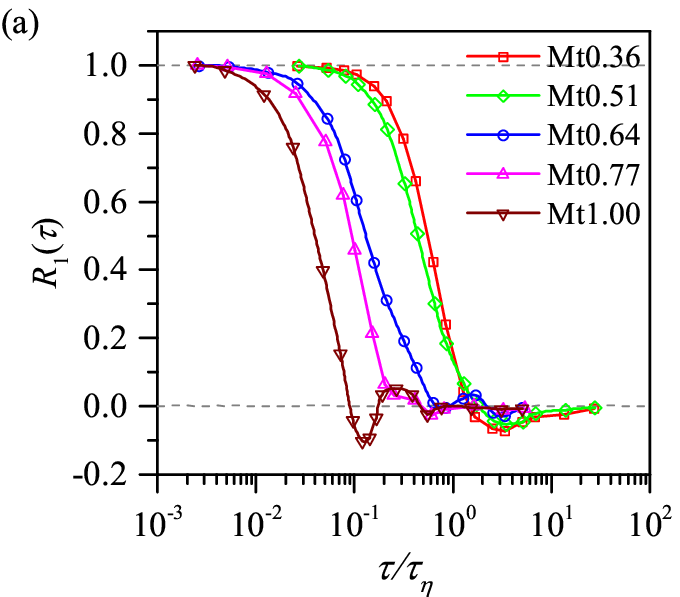}
  \caption{The autocorrelation function, $R_1(\tau)$, of $a_1$ at different $M_t$. (a) is for tracers; (b) is for light particles.}
\label{fig:Autocorltn}
\end{figure}

\begin{figure}
\centering
  \includegraphics[width=0.49\textwidth]{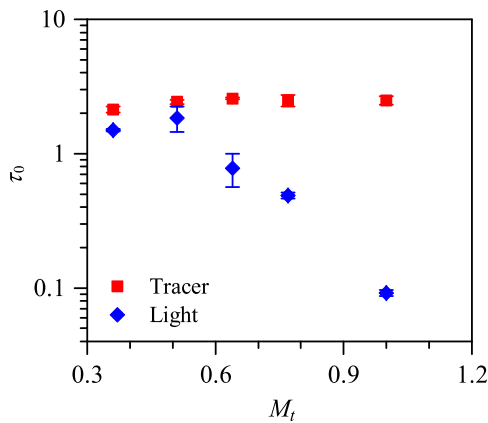}
  \caption{$\tau_0$ varies with $M_t$ for tracers (red squares) and light particles (blue diamonds).}
\label{fig:tau_0}
\end{figure}

Figure \ref{fig:Autocorltn} shows the autocorrelation function
of first component of the acceleration
{$R_1(\tau)=\langle a_1(t)a_1(t+\tau)
\rangle / \langle a_1(t)^2 \rangle$}
(because of isotropy $R_2$ and $R_3$ would be equal to $R_1$).
Ones can notice that for tracer particles,
the zero-crossing time ($\tau_0$) barely varies with $M_t$
and it is approximately close to $2.3\tau_{\eta}$,
which is similar to the results in incompressible turbulence
\citep{yeung1997one, mordant2004three, volk2008acceleration}.
Here, $\tau_0$ denotes the time in which $\rho_1(\tau)$ decreases to 0.
For light particles, $\tau_0$ is also approximate to
2.3 $\tau_{\eta}$ when $M_t < 0.6$.
However, as $M_t$ increases further, $\tau_0$ gradually decreases,
as shown in Fig. \ref{fig:tau_0}.

\begin{figure}
\centering
  \includegraphics[width=0.48\textwidth]{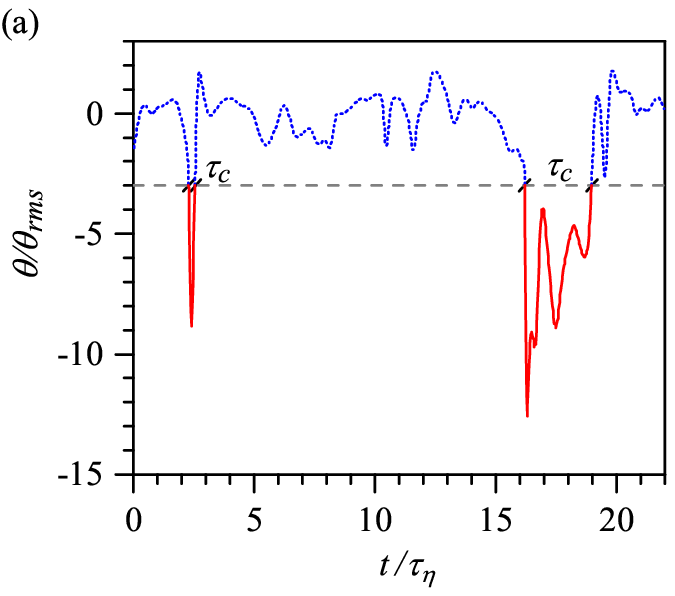}
  \includegraphics[width=0.48\textwidth]{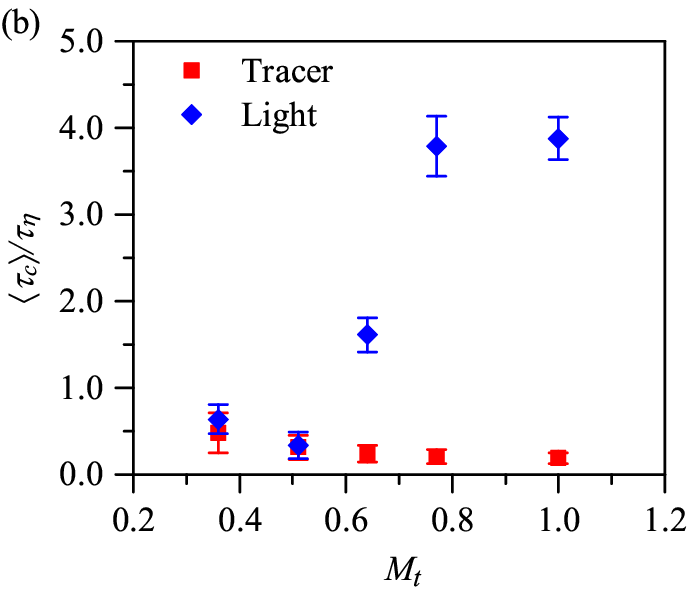}
  \caption{Panel (a): the history curve of $\theta$ along one trajectory of light particles at $M_t=1.00$. Here $\tau_c$ denotes the time over which particles pass through CR3.
  Panel (b): the average value of characteristic time of particles
  staying in CR3, $\langle \tau_c \rangle$, at different $M_t$.}
  \label{fig:tau_c_time}
\end{figure}

\begin{figure}
\centering
\centering
  \includegraphics[width=0.49\textwidth]{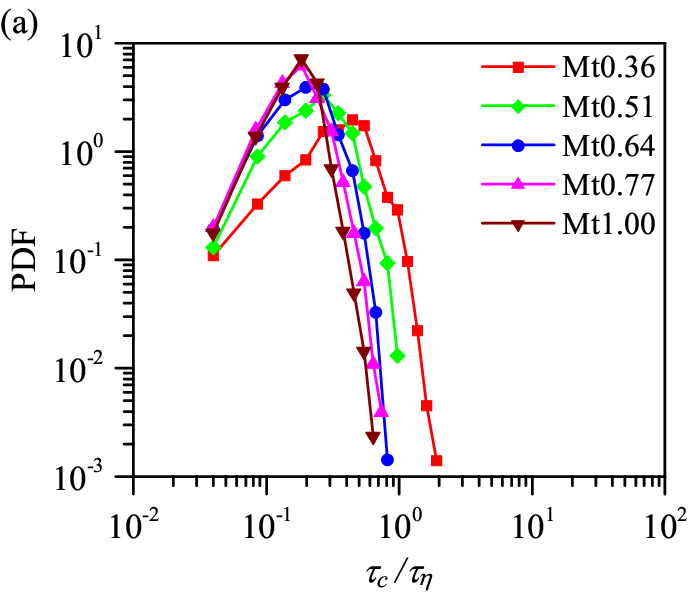}
  \includegraphics[width=0.49\textwidth]{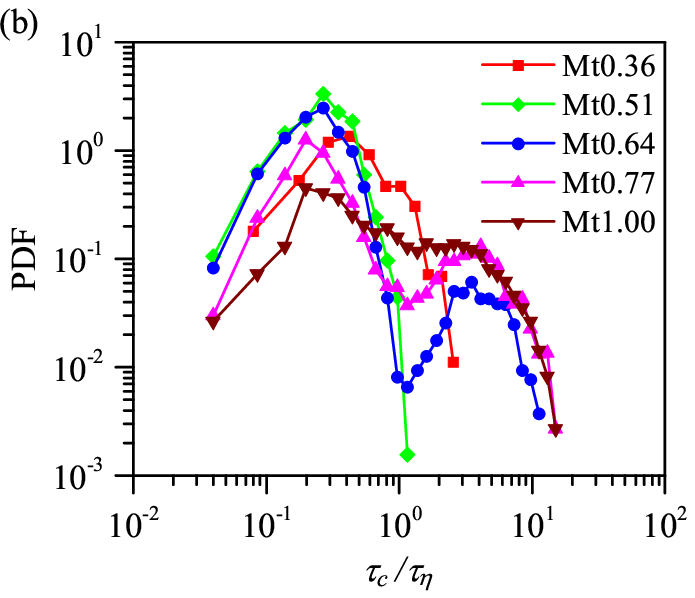}
  \caption{The PDF of the time ($\tau_c$)
  that particles stay in strong compression regions (CR3) at different $M_t$.
  (a) is for tracers; (b) is for light particles}
\label{fig:tau_c_PDF}
\end{figure}

Why is the behavior of $\tau_0$
between tracer and light particles different from each other?
\citet{yang2013acceleration} found
that the autocorrelation function of acceleration of tracer particles
near shocklets decreases much faster
than that near vortices in CHIT.
In addition, we use $\tau_c$ to denote the time over which particles pass through the strong compression regions (CR3), as shown in Fig. \ref{fig:tau_c_time} (a).
It is found that the average of $\tau_c$ among tracer particles,
$\langle \tau_c \rangle$, is distinctly smaller than $\tau_{\eta}$
and it slightly decreases with the increase of $M_t$
($\langle \tau_c \rangle \approx 0.2\tau_{\eta}$
at $M_t = 1.00$), as shown in Fig. \ref{fig:tau_c_time} (b).
For light particles, $\langle \tau_c \rangle$
is similar to that of tracer particles when {$M_t < 0.6$}.
However, as $M_t$ further increases,
$\langle \tau_c \rangle$ increases dramatically,
up to around 4$\tau_{\eta}$ at high $M_t$,
Then, it can be speculated that
$R_1(\tau)$ is barely influenced by shocklets
since the time that tracer particles
stay in strong compression regions (near shocklets) is quite short,
so that $\rho_1(\tau)$ should be mainly influenced by vortices.
Therefore, $\tau_0\approx 2.34\tau_{\eta}$ of tracer particles
is independent of $M_t$ and close to the small eddy turnover time.
For light particles, at low $M_t$ (e.g. $M_t=0.51$),
$\langle \tau_c \rangle$ is similar to that of tracers
so that $\tau_0$ of light particles is quite close to
that of tracer particles.
However, as $M_t$ increases,
$\langle \tau_c \rangle$ of light particles increases gradually
because more light particles will be trapped
in strong compression regions (near shocklets)
for several $\tau_{\eta}$, as shown in Fig. \ref{fig:tau_c_PDF} (b)
where the PDFs of $\tau_c$ of light particles
present two peaks when $M_t > 0.6$,
{at difference from what happens for tracers
(seen in Fig. \ref{fig:tau_c_PDF} (a))}.
The  position of first peak ($\tau_c \slash \tau_{\eta}=0.2-0.3$)
is similar to that of tracer particles.
However, the second peak occurs in the range of
$\tau_c \slash \tau_{\eta}=3.0-4.0$,
indicating that many light particles are trapped by shocklets,
hence, the influence  on $R_1(\tau)$ becomes remarkable.
Therefore, $\tau_0$ will be influenced
by both shocklets and vortices at high $M_t$,
and as $M_t$ increases, $\tau_0$ of light particles decreases,
as shown in Fig. \ref{fig:Autocorltn}(b),
since more light particles are trapped near shocklets.

\section{Conclusion and Discussion}\label{sec:ConclusionandDiscussion}
The acceleration statistics of tracer and light particles
in CHIT have been studied.
Our main finding is related to the characteristic signature
of the presence of shocklets in the probability distribution function
of acceleration and on its characteristic correlation times.
In particular, we found that at $ M_t \sim 0.6$ the statistical properties
of light particles acceleration start to be strongly different from
the one of the underlying tracers,
developing a sort of {\it condensation} on and near shocklets structures,
where a high percentage of the total number of particles is concentrated.
As a consequence, acceleration flatness,
skewness and autocorrelation time are strongly affected.
We find that in the range here investigated $M_t \in [0.31,1]$
the flatness factor of acceleration components, $F_a$,
of tracers increases monotonically with $M_t$.
For light particles, $F_a$ also increases with $M_t$ at $M_t< 0.6$.
However, when $M_t>0.6$, $F_a$ of light particle decreases with $M_t$
because of preferential accumulation in strong compression regions.

The PDFs of the longitudinal acceleration of tracer particles
are found to be skewed towards negative value,
and as $M_t$ increases, the skewness factor, $S_a$, becomes more negative.
For light particles, $S_a$ also becomes more negative
with the increase of $M_t$,
then followed by a tendency to return to 0 after $M_t > 0.6$.
We attribute the tendency of the longitudinal acceleration PDF
of light particles to become more and more symmetric at high $M_t$
to the fact that more particles reverse their direction of velocity
from $\alpha < 90^{\circ}$ to $\alpha > 90^{\circ}$ in compression regions
(e.g. the regions with $\theta < -\theta_{rms}$ ) at increasing compressibility. Together with the fact that more light particles with positive longitudinal acceleration enter compression regions at higher $M_t$.

For tracers, when the normalized transverse acceleration magnitude,
$\widetilde{a}_T$, is not large (e.g. $\widetilde{a}_T<3$),
the PDFs of $\widetilde{a}_T$ are close to each other for different $M_t$.
This implies that when $\widetilde{a}_T$ is small,
$\widetilde{a}_T$ is mainly influenced by vortices, rather than shocklets.
On the contrast, for light particles,
the PDFs of $\widetilde{a}_T$ are quite distinguishing for different $M_t$.
At higher $M_t$, the probability of small $\widetilde{a}_T$ is much larger,
indicating that $\widetilde{a}_T$ of light particles is easier to be influenced
by shocklets than that of tracer particles.

We have found that for tracer particles,
the characteristic time, $\tau_c$,  near shocklets is quite short,
especially at high $M_t$
($\langle \tau_c \rangle \approx 0.2\tau_{\eta}$ at $M_t=1.00$).
Then, the autocorrelation function of accelerations
is barely influenced by shocklets.
As a result, the zero-crossing time, $\tau_0$,
does not vary too much with $M_t$.
For light particles, as $M_t$ increases,
more particles are trapped near shocklets for several $\tau_{\eta}$
so that the influence of shocklets on the autocorrelation function
of accelerations becomes remarkable.
Therefore, $\tau_0$ is greatly influenced by shocklets and decreases with $M_t$.

The strong singular signature of shocklets for the dynamic
and statistics of light particles opens many important questions
that needs further numerical and experimental studies.
In particular, the role of correction terms to the Eq. (\ref{eq:dv_dt})
induced by the presence of compressibility must be better elucidated
as discussed in Appendix A.
Second, in the presence of strong particles' concentration nearby shocklets,
the importance of bubble-bubble collision, breakup and coalescence
should be better quantified, as well as their feedback on the flow.

We are grateful to Prof. Jianchun Wang for providing us the DNS code for CHIT.
This work is funded by the National Numerical Wind Tunnel
Project (No. NNW2019ZT1-A01), the National Natural Science Foundation of China
(Grant No. 91752201), the Department of Science and Technology of Guangdong Province
(Grant No. 2019B21203001), Key Special Project for Introduced Talents Team
of Southern Marine Science and Engineering Guangdong Laboratory (Guangzhou)
(Grant No. GML2019ZD0103), and by the Shenzhen Science and Technology Innovation Committee
(Grant No. KQTD20180411143441009).
We acknowledge computing support provided by the Center for Computational Science
and Engineering of the Southern University of Science and Technology.
L.B. acknowledges hospitality from  Southern University of Science and Technology and  funding from the European Research Council (ERC) under the European Unions Horizon 2020 research and innovation programme (grant agreement No 882340).

Declaration of interests. The authors report no conflict of interest.

\appendix
\section{}\label{AppdxA}
\begin{figure}
\centering
  \includegraphics[width=0.49\textwidth]{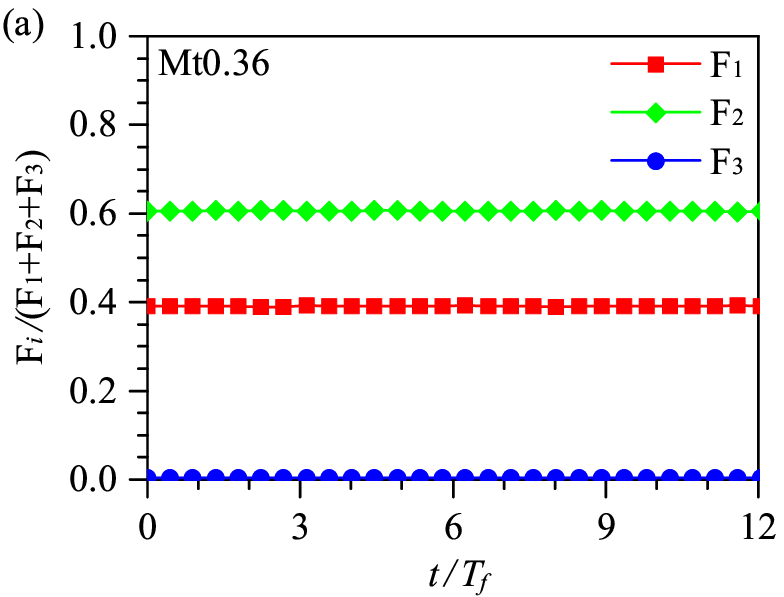}
  \includegraphics[width=0.49\textwidth]{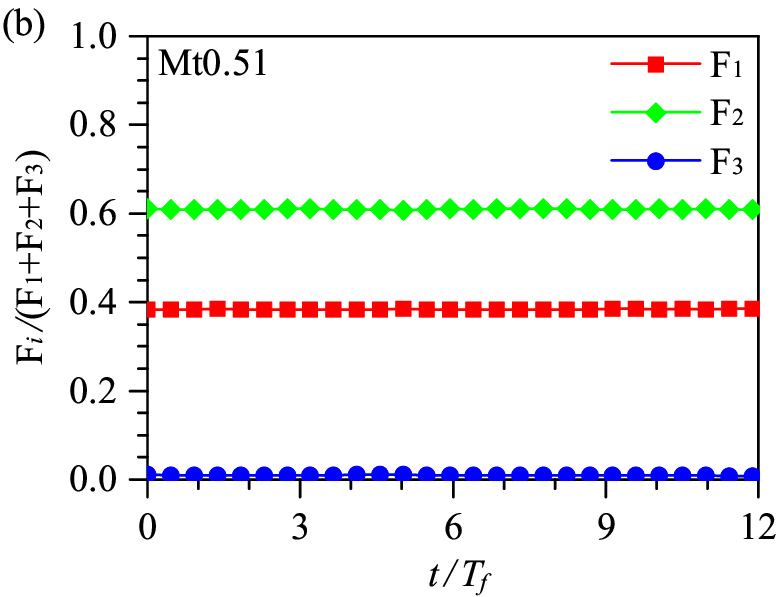}

  \includegraphics[width=0.49\textwidth]{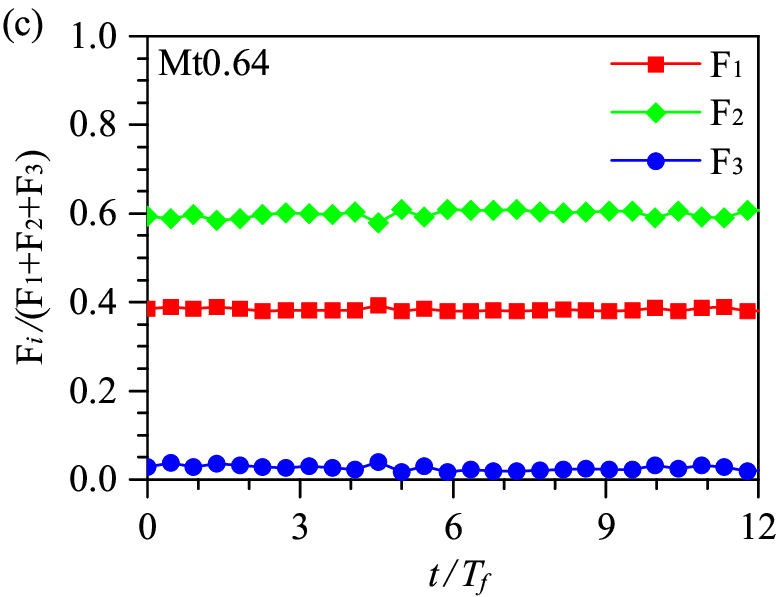}
  \includegraphics[width=0.49\textwidth]{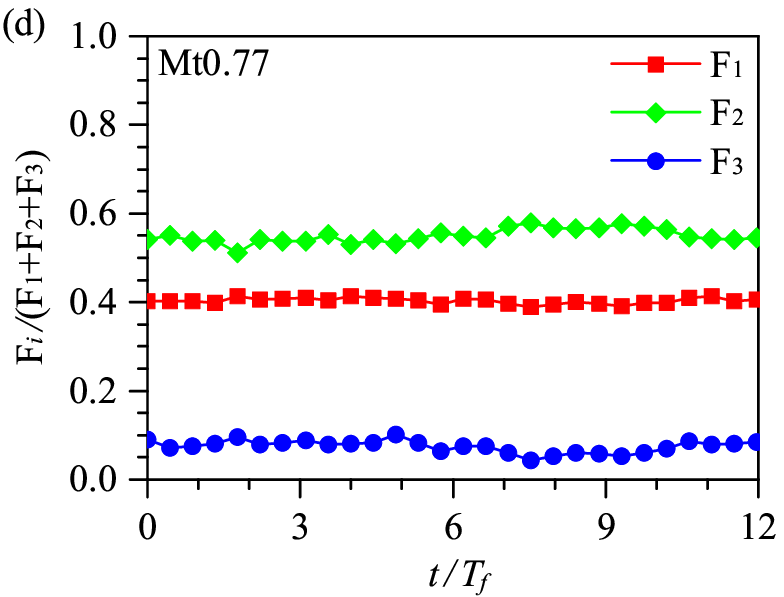}

  \includegraphics[width=0.49\textwidth]{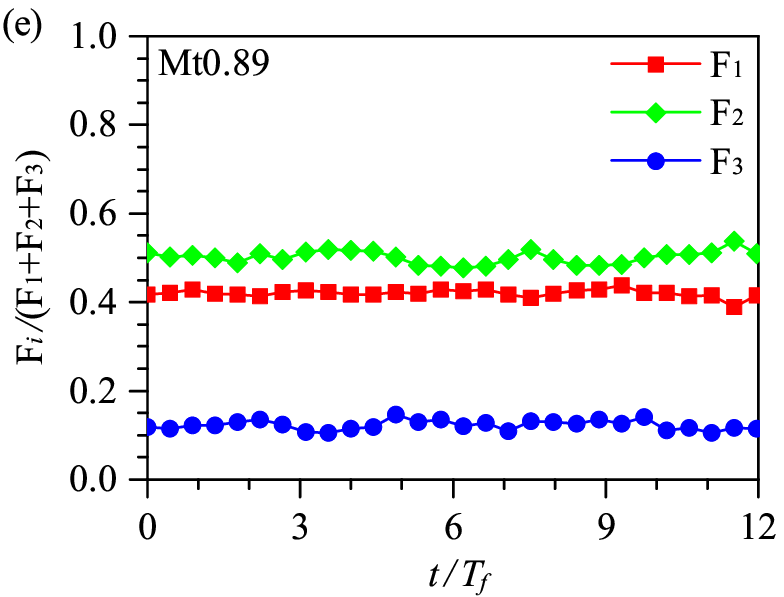}
  \includegraphics[width=0.49\textwidth]{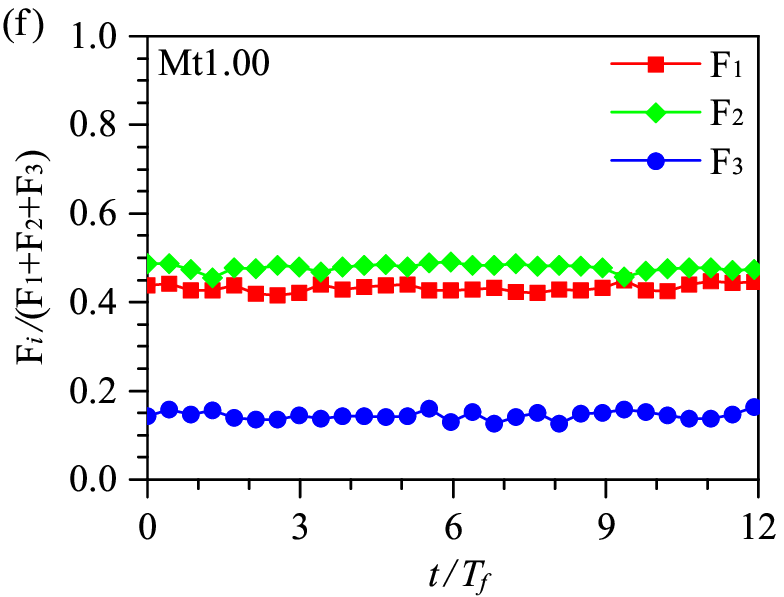}
  \caption{The average values of $\mathrm{F_1}$, $\mathrm{F_2}$ and $\mathrm{F_3}$ among all particles at different time slices}
\label{figA:appendix}
\end{figure}
According to the work \citep{parmar2012equation},
the governing equation for the movement of light particles
in compressible turbulence can be expressed as below
(neglecting history force):
\begin{equation}
  \rho_p \frac{\mathrm{d} \textbf{v}_p}{\mathrm{d} t}=\frac{9\mu}{2a^2}
  \left(\textbf{\emph{u}}_p-\textbf{v}_p \right)+\rho_f\frac{\mathrm{D} \textbf{\emph{u}}_p}{\mathrm{D} t}+\frac{1}{2}\left[\frac{\mathrm{D} \left(\rho_f \textbf{\emph{u}}_p \right)}{\mathrm{D} t}-\frac{\mathrm{d} \left(\rho_f \textbf{v}_p \right)}{\mathrm{d} t}\right]
  \label{EqA:dv_dt1}
\end{equation}
\begin{equation}
  \frac{\mathrm{d} \textbf{v}_p}{\mathrm{d} t}=\frac{\textbf{\emph{u}}_p-\textbf{v}_p}{\tau_p}+\beta\frac{\mathrm{D} \textbf{\emph{u}}_p}{\mathrm{D} t} +\frac{\beta}{3\rho_f}\left[\textbf{\emph{u}}_p\frac{\mathrm{D} \rho_f}{\mathrm{D} t}-\textbf{v}_p\frac{\mathrm{d} \rho_f}{\mathrm{d} t}\right]
  \label{EqA:dv_dt2}
\end{equation}
from mass conservation equation, we have
\begin{equation}
  \frac{\mathrm{D}\rho_f}{\mathrm{D}t}=-\rho_f \nabla \cdot \textbf{\emph{u}}
  \label{EqA:DuDt}
\end{equation}
\begin{equation}
  \frac{\mathrm{d}\rho_f}{\mathrm{d}t}=\frac{\mathrm{D}\rho_f}{\mathrm{D}t}
  +\left(\textbf{v}_p-\textbf{\emph{u}}_p \right)\cdot \nabla \rho_f
  \label{EqA:dudt}
\end{equation}
substituting Eq. (\ref{EqA:DuDt}) and Eq. (\ref{EqA:dudt})
into Eq. (\ref{EqA:dv_dt2}), we get
\begin{equation}
  \frac{\mathrm{d} \textbf{v}_p}{\mathrm{d} t}=\frac{\textbf{\emph{u}}_p-\textbf{v}_p}{\tau_p}+\beta\frac{\mathrm{D} \textbf{\emph{u}}_p}{\mathrm{D}t}+\left\{ \frac{\beta}{3}\left(\textbf{v}_p-\textbf{\emph{u}}_p\right)\nabla \cdot \textbf{\emph{u}}_p-\frac{\beta}{3\rho_f}\textbf{v}_p\left[\left(\textbf{v}_p
  -\textbf{\emph{u}}_p\right)\cdot\nabla \rho_f\right]\right\}
  \label{EqA:dv_dt3}
\end{equation}

If St is small ($\tau_p$ is quite small),
($\textbf{v}_p-\textbf{\emph{u}}_p$) in the third term
at the right hand side (RHS) of Eq. (\ref{EqA:dv_dt3}) will be also small
as $\textbf{v}_p=\textbf{\emph{u}}_p+O(\tau_p)$.
However, in compressible flows, $\nabla \cdot \textbf{\emph{u}}_p$
and $\nabla \rho_f$ will increase  as $M_t$ increases.
Consequently, there is a question whether the contribution of the third term
at RHS of Eq. (\ref{EqA:dv_dt3}) can be neglected.
Here, we define
  $$\mathrm{F_1}=\Big|\frac{\textbf{\emph{u}}_p-\textbf{v}_p}{\tau_p}\Big|$$
  $$\mathrm{F_2}=\Big|\beta\frac{\mathrm{D} \textbf{\emph{u}}_p}{\mathrm{D}t}\Big|$$
  $$\mathrm{F_3}=\Big|\frac{\beta}{3}\left(\textbf{v}_p-\textbf{\emph{u}}_p\right)
  \nabla \cdot\textbf{\emph{u}}_p-\frac{\beta}{3\rho_f}\textbf{v}_p \left[\left(\textbf{v}_p-\textbf{\emph{u}}_p\right)\cdot\nabla\rho_f\right]\Big|$$
and then measure the averages of $\mathrm{F_1}$,
$\mathrm{F_2}$ and $\mathrm{F_3}$ among all particles
at different time slices, in order to have an apriori estimate
of the importance of the neglected terms.
In Fig. \ref{figA:appendix} we show that the contribution of $\mathrm{F_3}$
is well below $5\%$ up to $M_t=0.6$ and becomes of the order of $10-15\%$
only at the largest Mach number $M_t=1.00$.
As a result, we argue that the approximation we made is reasonably acceptable
up to the largest Mach number we have investigated and
we leave for future studies the question to check the precise impact
of the extra terms in (\ref{EqA:dv_dt3}) for those Mach regimes.


\bibliographystyle{jfm}
\bibliography{jfm-instructions}

\end{document}